\documentclass[12pt]{article}

\usepackage{amsfonts}
\usepackage[margin=1in,nohead,a4paper,dvips]{geometry} 

\makeatletter
\let\@xp=\expandafter
\let\@nx=\noexpand
\providecommand\numberwithin[2]{%
  \@ifundefined{c@#1}{\@nocounterr{#1}}{%
    \@ifundefined{c@#2}{\@nocounterr{#2}}{%
    \@addtoreset{#1}{#2}%
    \toks@\@xp\@xp\@xp{\csname the#1\endcsname}%
    \@xp\xdef\csname the#1\endcsname
      {\@xp\@nx\csname the#2\endcsname
       .\the\toks@}}}} 
\def\eqnarray{%
   \stepcounter{equation}%
   \def\@currentlabel{\p@equation\theequation}%
   \global\@eqnswtrue
   \m@th
   \global\@eqcnt\z@
   \tabskip\@centering
   \let\\\@eqncr
   $$\everycr{}\halign to\displaywidth\bgroup
       \hskip\@centering$\displaystyle\tabskip\z@skip{##}$\@eqnsel
      &\global\@eqcnt\@ne\hfil$\displaystyle{\hbox{}##\hbox{}}$\hfil
      &\global\@eqcnt\tw@ $\displaystyle{##}$\hfil\tabskip\@centering
      &\global\@eqcnt\thr@@ \hb@xt@\z@\bgroup\hss##\egroup
         \tabskip\z@skip
      \cr
}
\def\mbar#1{\kern 0.1em\overline{\kern -0.1em #1 \kern -0.1em} 
  \kern 0.1em}

\def\cJ#1{\mathcal J^{#1}}
\def\cW{\mathcal W}
\def\bcJ#1{\bar \mathcal J^{#1}}
\def\bSigma {\kern 0.25em\overline{\kern -0.25em \Sigma \kern -0.1em}
\kern 0.1em}
\def\bW {\kern 0.1em\overline{\kern -0.1em W \kern -0.1em} \kern
0.1em}
\def\bcW {\kern 0.1em\overline{\kern -0.1em \mathcal W \kern -0.1em}
\kern 0.1em}
\def\bX {\kern 0.25em\overline{\kern -0.25em X \kern -0.1em} \kern
0.1em}

\def\Ad {\mathop {\operator@font Ad}\nolimits}
\def\id {\mathop {\operator@font id}\nolimits}
\def\Mat {\mathop {\operator@font Mat}\nolimits}
\def\tr {\mathop {\operator@font tr}\nolimits}
\makeatother

\numberwithin{equation}{section} 

\def\bbC{\mathbb C}

\def\bbR{\mathbb R}
\def\bbZ{\mathbb Z}
\def\cmotms{\mathop{\, \otimes \,}\limits_{\raise .4em \hbox{,}}}
\def\cW{\mathcal W}
\def\gothg{\mathfrak g}

\def\dif {\mathrm d \kern .1em}

\mathchardef\Gamma="100
\mathchardef\Lambda="103
\mathchardef\Phi="108
\mathchardef\Sigma="106

\begin{document}

\title{$W$-algebras for non-abelian Toda systems}
\author{Kh. S. Nirov\\
\small Institute for Nuclear Research of the Russian Academy of
Sciences\\[-.3em]
\small 60th October Anniversary Prospect 7a, 117312 Moscow,
Russia\\[-.3em]
\small E-mail: nirov@ms2.inr.ac.ru\\[.3em]
A. V. Razumov\\
\small Institute for High Energy Physics\\[-.3em]
\small 142280 Protvino, Moscow Region, Russia\\[-.3em]
\small E-mail: razumov@mx.ihep.su}
\date{}

\maketitle

\begin{abstract}
We construct the classical $W$-algebras for some non-abelian Toda
systems associated with the Lie groups $\mathrm{GL}_{\, 2n}(\bbR)$
and $\mathrm{Sp}_n(\bbR)$. We start with the set of characteristic
integrals and find the Poisson brackets for the corresponding
Hamiltonian counterparts. The convenient block matrix representation
for the Toda equations is used. The infinitesimal symmetry
transformations generated by the elements of the $W$-algebras 
are presented.
\end{abstract}

\section{Introduction}

The Toda systems constitute a remarkable class of two-dimensional
integrable systems. According to the group-algebraic approach
\cite{LSa92,RSa97} such a system is specified by the choice of 
a Lie group $G$ whose Lie algebra $\gothg$ is endowed with a
$\bbZ$-gradation. There exist so-called higher grading 
\cite{Lez95,GSa95} and multi-dimensional \cite{RSa97a,RSa97b}
generalizations of the Toda systems.

The `space-time' for a Toda system is a two-dimensional manifold, and
the `field space' is the Lie group $G_0$ corresponding to the Lie
subalgebra $\gothg_0$ of $\gothg$ corresponding to zero value of the
grading index. If the group $G_0$ is abelian the corresponding Toda
system is said to be abelian, otherwise one has a non-abelian Toda
system. There is a lot of papers devoted to abelian Toda systems,
while non-abelian Toda systems are not very well studied yet. This is
connected to the fact that until recently there was no a convenient
representation for such systems. It was shown in paper \cite{RSa97c}
that some class of non-abelian Toda systems can be represented in a
simple block matrix form. Later it was proven that it is the case 
for all Toda systems associated with classical semisimple Lie
groups~\cite{RSZ99}. This led to the renewal of the interest to this
class of integrable systems; see, for example, papers 
\cite{EGR97,EGR98,Lez98,Lez99}.

In the present paper we investigate the symmetries of the simplest
non-abelian Toda systems associated with finite dimensional Lie
groups $\mathrm{GL}_{\, 2n}(\bbR)$ and $\mathrm{Sp}_n(\bbR)$. Actually
for the systems under consideration there are an evident symmetry
resembling the symmetry of a Wess--Zumino--Novikov--Witten (WZNW)
model \cite{Nov82,Wit84} and the conformal symmetry. These symmetries
do not exhaust all symmetries of the systems. More symmetries can be
found using the so-called characteristic integrals whose existence is
related to the integrability of Toda systems. These integrals give an
infinite set of the densities of conserved charges. The Hamiltonian
counterparts of these conserved charges generate the required symmetry
transformations.

Thus, our strategy is as follows. We find the characteristic
integrals for our systems. Then we proceed to the Hamiltonian
formalism and find the Hamiltonian counterparts of the characteristic
integrals and conserved charges. This allows us to find the form of
infinitesimal symmetry transformations in the Hamiltonian formalism
and write down their Lagrangian version.

We show also that the set of characteristic integrals is closed with
respect to the Poisson bracket and form an object usually called a
classical $W$-algebra. The distinctive features of such algebras is
that their defining relations are essentially nonlinear and that they
contain Virasoro algebras corresponding to the conformal invariance.
The systematic study of $W$-algebras in the framework of general
quantum conformal field theory was initiated by A.~B.~Zamolodchikov
\cite{Zam85}. For a detailed review of the subject we refer the
reader to paper \cite{BSc93}.

Although our paper contains original results, in some parts it has
character of a review. It is worth to note here that majority of the
results on $W$-algebras for Toda systems was obtained by the method of
Hamiltonian reduction that is based on the fact that Toda systems
can be obtained if one starts with a WZNW model based on a Lie group
$G$ and then imposes relevant constraints on the conserved currents
forming with respect to the Poisson bracket two copies of loop
algebras associated with the Lie algebra $\gothg$
\cite{BFORFW90,FORTW92a,FORTW92b}. Here the Toda `field space' arises
as a factor in the generalized Gauss decomposition of the Lie group
$G$, which is valid only for a dense subset of $G$. This results in
that the true reduced system is different from a Toda system; see, 
in this respect, papers
\cite{TFe95,Ful96,FTs97,KTs96,RYa97,BFP98,BNVW99}. Such our 
conclusion is justified at least by the fact that the Toda systems
have singular solutions corresponding to some nonsingular initial
conditions, and that is impossible for a system being a reduction 
of a WZNW model which does not have such solutions. Thus, the results 
on Toda systems obtained with the help of the method of Hamiltonian
reduction require verification. 

It seems to us that the direct method used in our paper is more
appropriate to the problem under consideration than the method of
Hamiltonian reduction. In particular, it allows to identify the
generators of the Virasoro algebras describing the conformal
properties of the model with the Hamiltonian counterparts of the
components of the conformally improved energy-momentum tensor 
which is constructed by a standard procedure.

\section{Toda systems} \label{te}

In accordance with the group-algebraic approach \cite{LSa92,RSa97}
the construction of equations describing a Toda system looks as
follows. Let $G$ be a real or complex Lie group whose Lie algebra
$\gothg$ is endowed with a $\mathbb Z$-gradation,
\[
\gothg = \bigoplus_{m \in \mathbb Z} \gothg_m, \qquad
[\gothg_m,\gothg_n] \subset \gothg_{m+n}.
\]
Recall that for a given $\mathbb Z$-gradation of $\gothg$ the
subspace
$\gothg_0$ is a subalgebra of $\gothg$. The subspaces
\[
\gothg_{<0} = \bigoplus_{m < 0} {\gothg}_m, \qquad \gothg_{>0} =
\bigoplus_{m > 0} {\gothg}_m
\]
are also subalgebras of $\gothg$. Denote by $G_0$, $G_{<0}$ and
$G_{>0}$ the connected Lie subgroups of $G$ corresponding to the
subalgebras $\gothg_0$, and $\gothg_{<0}$ and $\gothg_{>0}$
respectively.

Let $M$ be a real two-dimensional manifold. Introduce local
coordinates on $M$ and denote them by $z^-$ and
$z^+$. One can also consider the case when $M$ is a one
dimensional complex manifold. In this case $z^-$ is a complex
coordinate on $M$ and $z^+$ is the complex conjugate of $z^-$. Let
$a_-$ and $a_+$ be some fixed mappings from $M$ to $\gothg_{-1}$ and
$\gothg_{+1}$, respectively, satisfying the relations
\begin{equation}
\partial_+ a_- = 0, \qquad \partial_- a_+ = 0. \label{6}
\end{equation}
Here and below we denote the partial derivatives over $z^-$ and $z^+$
by $\partial_-$ and $\partial_+$. Actually we assume that the
subspaces $\gothg_{-1}$ and $\gothg_{+1}$ are nontrivial. Generally,
if $l$ is a positive integer such that the subspaces $\gothg_m$ are
trivial for $-l < m <0$ and $0 < m < l$, one defines $a_-$ and $a_+$
as mappings from $M$ to $\gothg_{-l}$ and $\gothg_{+l}$ respectively.
Restrict ourselves to the case when $G$ is a matrix Lie group.
In other words, assume that for some positive integer $N$ it is a Lie
subgroup of the Lie group GL$_N(\bbR)$ or of the Lie group
GL$_N(\bbC)$. More general case is discussed in paper \cite{RSa94}.
In the case under consideration the equations describing the Toda
system are matrix partial differential equations of the form 
\begin{equation}
\partial_+(\gamma^{-1} \partial_- \gamma) 
= [a_-, \gamma^{-1} a_+ \gamma], \label{1}
\end{equation}
where $\gamma$ is a mapping from $M$ to $G_0$. Note that the 
equations (\ref{1}) can also be written as
\begin{equation}
\partial_-(\partial_+ \gamma \gamma^{-1}) 
= [\gamma a_- \gamma^{-1}, a_+]. \label{9}
\end{equation}
Parametrizing the group $G_0$ by a set of independent parameters,
or, in other words, introducing some coordinates on $G_0$, we can
rewrite the Toda equations as a system of equations for ordinary
functions, which we call Toda fields.

If the Lie group $G_0$ is abelian we say that we deal with an 
abelian Toda system, otherwise we call the system a non-abelian 
one. The complete classification of the Toda systems associated 
with the classical Lie groups is given in paper \cite{RSZ99}.

There is a constructive procedure of obtaining the general solution
to Toda equations \cite{RSa97,RSa97a,RSa97b}. It is based on the use
of the Gauss decomposition related to the $\mathbb Z$-gradation under
consideration. Here the Gauss decomposition is the representation of
an element of the Lie group $G$ as a product of elements of the
subgroups $G_{<0}$, $G_{>0}$ and $G_0$ taken in an appropriate order.
Another approach is based on the theory of representations of Lie
groups \cite{LSa92,RSa97}. 

In this paper we consider the simplest examples of non-abelian Toda
equations based on the Lie groups $\mathrm{GL}_{\, 2n}(\bbR)$ and
$\mathrm{Sp}_n(\bbR)$ \cite{OWi90,RSa97b}. 

We start with the Lie group $G = \mathrm{GL}_{\, 2n}(\bbR)$. The case
of the Lie group $\mathrm{Sp}_n(\bbR)$ will be considered in section
\ref{nle}. The Lie algebra $\gothg = \mathfrak{gl}_{\, 2n}(\bbR)$ of
$\mathrm{GL}_{\, 2n}(\bbR)$ is formed by all real $2n \times 2n$
matrices. Below we represent an arbitrary $2n \times 2n$ matrix $x$ in
the block matrix form
\begin{equation}
x = \left( \begin{array}{cc}
x_{11} & x_{12} \\
x_{21} & x_{22}
\end{array} \right), \label{56}
\end{equation}
where $x_{rs}$, $r,s = 1,2$, are $n \times n$ matrices.

Recall that an element $q \in \gothg$ is said to be the grading
operator generating the $\bbZ$-gradation under consideration if
\[
\gothg_m = \{ x \in \gothg \mid [q, x] = m x \}.
\]
In particular, any $\bbZ$-gradation of a finite dimensional complex
semisimple Lie algebra is generated by the corresponding grading
operator.

Denote by $I_n$ the unit $n \times n$ matrix. It is easy to show that
the element
\begin{equation}
q = \frac{1}{2} \left( \begin{array}{cc}
I_n & 0 \\
0 & - I_n
\end{array} \right) \label{13}
\end{equation}
generates a $\bbZ$-gradation of $\mathfrak{gl}_{\, 2n}(\bbR)$. Here
the subspaces $\gothg_{-1}$ and $\gothg_{+1}$ are the sets formed by
all block strictly lower triangular and strictly upper triangular
matrices of $\mathfrak{gl}_{\, 2n}(\bbR)$, respectively, and the
subspace $\gothg_0$ is the set of all block diagonal matrices of
$\mathfrak{gl}_{\, 2n}(\bbR)$. All other grading subspaces are
trivial, and we have $\gothg_{<0} = \gothg_{-1}$, $\gothg_{>0} =
\gothg_{+1}$.

Hence, the general form of the mappings $a_-$ and $a_+$ is
\begin{equation}
a_- = \left( \begin{array}{cc}
0 & 0 \\
A_- & 0
\end{array} \right), \qquad
a_+ = \left( \begin{array}{cc}
0 & A_+ \\
0 & 0
\end{array} \right), \label{57}
\end{equation}
where $A_-$ and $A_+$ are arbitrary $n \times n$ matrix-valued
functions on $M$ satisfying the condition
\begin{equation}
\partial_+ A_- = 0, \qquad 
\partial_- A_+ = 0. \label{58}
\end{equation}
In this paper we restrict ourselves to the case 
$A_- = I_n$ and $A_+ = I_n$.

It is not difficult to describe the corresponding subgroups $G_{<0}$,
$G_{>0}$ and $G_0$ of the Lie group $\mathrm{GL}_{\, 2n}(\bbR)$. The
subgroups $G_{<0}$ and $G_{>0}$ consist of all block lower triangular
and upper triangular matrices of $\mathrm{GL}_{\, 2n}(\bbR)$,
respectively, with unit matrices on the diagonal. The subgroup $G_0$
is formed by all block diagonal matrices of $\mathrm{GL}_{\,
2n}(\bbR)$.

Parametrize the mapping $\gamma$ as
\begin{equation}
\gamma = \left( \begin{array}{cc}
\Gamma^{(1)} & 0 \\
0 & \Gamma^{(2)}
\end{array} \right), \label{26}
\end{equation}
where $\Gamma^{(1)}$ and $\Gamma^{(2)}$ are mappings from $M$ to
the Lie group $\mathrm{GL}_{\, n}(\bbR)$. With this
pa\-ra\-me\-tri\-zation we write the Toda equations in the form
\begin{equation}
\partial_+ (\Gamma^{(1)-1} \, \partial_- \Gamma^{(1)}) = -
\Gamma^{(1)-1} \, \Gamma^{(2)}, \qquad \partial_+
(\Gamma^{(2)-1} \, \partial_- \Gamma^{(2)}) = \Gamma^{(1)-1} \,
\Gamma^{(2)}, \label{36}
\end{equation}
or in the form
\begin{equation}
\partial_- (\partial_+ \Gamma^{(1)} \, \Gamma^{(1)-1}) = -
\Gamma^{(2)} \, \Gamma^{(1)-1}, \qquad \partial_-
(\partial_+\Gamma^{(2)} \, \Gamma^{(2)-1}) = \Gamma^{(2)} \,
\Gamma^{(1)-1}. \label{51}
\end{equation}
The exact general solution to these equations was obtained in
paper \cite{RSa97b}.

One can get convinced that the transformations
\begin{equation}
\Gamma^{(1)} \to \Lambda_+ \Gamma^{(1)} \Lambda^{}_-, \qquad
\Gamma^{(2)} \to \Lambda_+ \Gamma^{(2)} \Lambda^{}_-, \label{77}
\end{equation}
where $\Lambda_-$ and $\Lambda_+$ are mappings from $M$ to
$\mathrm{GL}_n(\bbR)$ satisfying the conditions
\[
\partial_+ \Lambda_- = 0, \qquad \partial_- \Lambda_+ = 0,
\]
are symmetry transformations for the system under consideration. This
symmetry, by an evident reason, can be called a WZNW-type symmetry.

The system possesses also the conformal symmetry. Here the conformal
transformations
\begin{equation}
z^- \to \zeta^-(z^-), \qquad z^+ \to \zeta^+(z^+) \label{78}
\end{equation}
act on the space of solutions of the equations (\ref{36}), 
or (\ref{51}), in the following way \cite{RSZ99}:
\begin{eqnarray}
& \Gamma^{(1)}(z^-, z^+) \to 
[\partial_- \zeta^-(z^-)\partial_+ \zeta^+(z^+)]^{-1/2} 
\Gamma^{(1)}(\zeta^+(z^+), \zeta^-(z^-)),
\label{71} \\[.3em]
& \Gamma^{(2)}(z^-, z^+) \to 
[\partial_- \zeta^-(z^-)\partial_+ \zeta^+(z^+)]^{1/2} 
\Gamma^{(2)}(\zeta^+(z^+), \zeta^-(z^-)).
\label{72}
\end{eqnarray}

The WZNW-type symmetry and the conformal symmetry do not exhaust all
symmetries of the system. To find additional symmetry transformations
we can use the following procedure. 

First we find conserved charges. In the case under consideration we
have an infinite set of conserved charges provided by the so-called
characteristic integrals. In the Hamiltonian formalism the conserved
charges generate symmetry transformations. So, we construct the
Lagrangian formulation for our system and then proceed to the
corresponding Hamiltonian description. After that we consider the
symmetry transformations generated by the Hamiltonian counterparts of
the conserved charges associated with the characteristic integrals,
and finally obtain their Lagrangian version. This allows us, in
particular, to obtain the WZNW-type symmetry transformations and
the conformal transformations discussed above.

\section{Characteristic integrals} \label{ci}

A characteristic integral of a Toda system is, by definition, either
a differential polynomial $W$ of the Toda fields satisfying the
relation
\begin{equation}
\partial_+ W = 0, \label{69}
\end{equation}
or a differential polynomial $\bW$ of the Toda fields which satisfy
the relation
\begin{equation}
\partial_- \bW = 0. \label{70}
\end{equation}
By a differential polynomial we mean a polynomial function of the
fields and their derivatives.

Let us treat the manifold $M$ as a flat Riemannian manifold with the
coordinates $z^-$ and $z^+$ being light-front coordinates and the
metric tensor $\eta$ having the form (\ref{A.33}). The usual flat
coordinates $z^0$ and $z^1$ are related to the light-front
coordinates $z^-$ and $z^+$ by the relation (\ref{A.34}). 
Using these coordinates we write the equality (\ref{69}) 
as
\[
\partial_0 W + \partial_1 W = 0,
\]
where $\partial_0 = \partial/\partial z^0$ and 
$\partial_1 = \partial/\partial z^1$. Hence, the function $W$ is 
a density of a conserved charge. Moreover, multiplying $W$ by a 
function which depends only on $z^-$ we again obtain a characteristic 
integral. Therefore, a characteristic integral generates an infinite 
set of densities of conserved charges. Similarly, multiplying a
characteristic integral satisfying the relation (\ref{70}) by 
functions depending only on $z^+$ we again obtain an infinite 
set of densities of conserved charges.

It is clear that any differential polynomial of characteristic
integrals is also a characteristic integral. Moreover, the Poisson
bracket of the Hamiltonian counterparts of any two characteristic
integrals is again a characteristic integral. Therefore, a necessary
step in investigation of characteristic integrals is to show that
they form a closed set with respect to the Poisson bracket, or, in
other words, that they form an object called a $W$-algebra; see, for
a review, \cite{BSc93}.

There are two main methods for obtaining characteristic integrals 
for Toda systems. The first one is based on the construction of a
generating pseudo-differential operator; see, for example, papers
\cite{LSa89,BGe89a,BGe89b,BFORFW90}. The second method is based on
the usage of the so-called Drinfeld--Sokolov gauge; see, for example,
papers \cite{DSo84,BFORFW90,FORTW92a,FORTW92b}. In the present
paper we use the latter method.

It is well known that Toda equations (\ref{1}) can be obtained as the
zero curvature condition for some connection on the trivial principal
fiber bundle $M \times G \to M$ \cite{LSa92,RSa97}. We identify the
connection under consideration with a $\gothg$-valued one-form 
$\omega$ on $M$. Using the basis formed by the 1-forms $\dif z^-$ 
and $\dif z^+$, we write
\[
\omega = \omega_- \dif z^- + \omega_+ \dif z^+,
\]
where the components $\omega_-$ and $\omega_+$ are $\gothg$-valued
functions on $M$. The curvature of the connection $\omega$ is zero if
and only if 
\begin{equation}
\dif \omega + \omega \wedge \omega = 0, \label{3}
\end{equation}
or, in terms of the components, 
\begin{equation}
\partial_- \omega_+ - \partial_+ \omega_- + [\omega_-, \omega_+] = 0.
\label{2}
\end{equation}
If we consider the components $\omega_-$ and $\omega_+$ of the form
\begin{equation}
\omega_- = a_- + \gamma^{-1} \partial_- \gamma, \qquad 
\omega_+ = \gamma^{-1} a_+ \gamma, \label{4}
\end{equation}
then the zero curvature condition (\ref{2}) is
equivalent to the Toda equations (\ref{1}).

Recall that the zero curvature condition is gauge invariant. It means
that if a connection $\omega$ satisfies the relation (\ref{3}), then 
for any mapping $\psi: M \to G$ the gauge transformed connection
\[
\omega^\psi = \psi^{-1} \omega \psi + \psi^{-1} \dif \psi
\]
satisfies the relation (\ref{3}) as well. In terms of the components 
one has
\[
\omega^\psi_- = \psi^{-1} \omega_- \psi + \psi^{-1} \partial_- \psi,
\qquad
\omega^\psi_+ = \psi^{-1} \omega_+ \psi + \psi^{-1} \partial_+ \psi.
\]
In particular, if we consider the connection with the components
given by (\ref{4}) and choose $\psi = \gamma^{-1}$ we will come to 
the connection, which we also denote by $\omega$, with the components
\begin{equation}
\omega_- = \gamma a_- \gamma^{-1} \qquad 
\omega_+ = - \partial_+\gamma \gamma^{-1} + a_+. \label{5}
\end{equation}
And the zero curvature condition (\ref{2}) gives the Toda equations
written in form (\ref{9}).

Let us return to our specific example of Toda equations. Write
the components $\omega_-$ and $\omega_+$ defined by (\ref{4}) in
the block matrix form
\[
\omega_- = \left( \begin{array}{cc}
\Sigma^{(1)}_- & 0 \\[.7em]
I_n & \Sigma^{(2)}_-
\end{array} \right), \qquad
\omega_+ = \left( \begin{array}{cc}
0 & \Gamma^{(1)-1} \Gamma^{(2)} \\[.7em]
0 & 0
\end{array} \right),
\]
where we denoted
\begin{equation}
\Sigma^{(1)}_- = \Gamma^{(1)-1} \partial_- \Gamma^{(1)}, \qquad
\Sigma^{(2)}_- = \Gamma^{(2)-1} \partial_- \Gamma^{(2)}. \label{14}
\end{equation}
Now consider the gauge transformation generated by a mapping 
$\psi: M \to G_{>0}$. The general form of such a mapping is
\[
\psi = \left( \begin{array}{cc}
I_n & \chi \\[.7em]
0 & I_n
\end{array} \right).
\]
For the component $\omega_-^\psi$ we obtain the expression
\[
\omega_-^\psi = \left( \begin{array}{cc}
\Sigma^{(1)}_- - \chi & (\Sigma^{(1)}_- - \chi) \, \chi - \chi \,
\Sigma^{(2)}_- + \partial_- \chi \\[.7em]
I_n & \Sigma^{(2)}_- + \chi
\end{array} \right).
\]
The Drinfeld--Sokolov gauge \cite{DSo84,BFORFW90,FORTW92a,FORTW92b}
in our case is
fixed
by the requirement
\[
(\omega_-^\psi)_{11} = (\omega_-^\psi)_{22}.
\]
It is clear that this requirement gives
\[
\chi = \frac{1}{2} \, ( \Sigma^{(1)}_- - \Sigma^{(2)}_- ),
\]
and we obtain
\[
\omega_-^\psi = \left( \begin{array}{cc}
\displaystyle - \frac{1}{2 \kappa} \, W_1 & \displaystyle -
\frac{1}{\kappa^2} W_2 + \frac{1}{4 \kappa^2} \, W_1^2 \\[.7em]
I_n & \displaystyle - \frac{1}{2 \kappa} \, W_1
\end{array} \right),
\]
where
\begin{equation}
W_1 = -\kappa (\Sigma^{(1)}_- + \Sigma^{(2)}_- ), \qquad
W_2 = - \kappa^2 \left( \frac{1}{2} \, (\partial_- \Sigma^{(1)}_- -
\partial_- \Sigma^{(2)}_-) - \Sigma^{(1)}_- \Sigma^{(2)}_- \right),
\label{16}
\end{equation}
and $\kappa$ is a constant. We introduced the constant $\kappa$ in
the
definition of the quantities $W_1$ and $W_2$ for future convenience.
Actually we will identify it with the constant entering the action of
the Toda theory.

For the component $\omega_+^\psi$ we have the expression
\[
\omega_+^\psi = \left( \begin{array}{cc}
0 & \displaystyle \Gamma^{(1)-1} \Gamma^{(2)} + \frac{1}{2} \,
(\partial_+ \Sigma^{(1)}_- - \partial_+ \Sigma^{(2)}_-) \\[.7em]
0 & 0
\end{array} \right).
\]
Therefore, if $\Gamma^{(1)}$ and $\Gamma^{(2)}$ satisfy the Toda
equations (\ref{36}) then $\omega_+^\psi = 0$, and the zero curvature
condition gives
\[
\partial_+ \omega_-^\psi = 0.
\]
This equality implies that
\begin{equation}
\partial_+ W_1 = 0, \qquad \partial_+ W_2 = 0. \label{19}
\end{equation}
Thus, the quantities $W_1$ and $W_2$ are matrix characteristic
integrals of the Toda system under consideration.

As we have already noted, any differential polynomial of
characteristic integrals is a characteristic integral. 
Therefore, they form a differential algebra. The generators of this 
algebra, the matrix elements of $W_1$ and $W_2$ in our case, can be 
chosen in different ways. Our choice is inspired by an intention to 
get simple expressions for Poisson brackets.

One can also start with the connection components of the form 
(\ref{5}). Performing the gauge transformation with
$\psi: M \to G_{<0}$,   
\[
\psi = \left( \begin{array}{cc}
I_n & 0 \\[.7em]
\displaystyle \frac{1}{2} (\bSigma^{(1)}_+ - \bSigma^{(2)}_+)
& I_n
\end{array} \right),
\]
where
\begin{equation}
\bSigma^{(1)}_+ = \partial_+ \Gamma^{(1)} \Gamma^{(1)-1}, \qquad
\bSigma^{(2)}_+ = \partial_+ \Gamma^{(2)} \Gamma^{(2)-1},
\label{15}
\end{equation}
we obtain
\[
\omega_+^\psi = \left( \begin{array}{cc}
\displaystyle \frac{1}{2 \kappa} \, \bW_1 & I_n \\[.7em]
\displaystyle - \frac{1}{\kappa^2} \bW_2 + \frac{1}{4 \kappa^2} \,
\bW_1^2 & \displaystyle \frac{1}{2 \kappa} \, \bW_1
\end{array} \right),
\]
where
\begin{equation}
\bW_1 = -\kappa (\bSigma^{(1)}_+ + \bSigma^{(2)}_+ ), \qquad \bW_2 =
-\kappa^2 \left( \frac{1}{2} \, (\partial_+ \bSigma^{(1)}_+ -
\partial_+ \bSigma^{(2)}_+) - \bSigma^{(2)}_+ \bSigma^{(1)}_+
\right).
\label{17}
\end{equation}
For the component $\omega_-^\psi$ one has
\[
\omega_-^\psi = \left( \begin{array}{cc}
0 & 0 \\[.7em]
\displaystyle \Gamma^{(2)} \Gamma^{(1)-1} + \frac{1}{2} \,
(\partial_- \bSigma^{(1)}_+ - \partial_- \bSigma^{(2)}_+) & 0
\end{array} \right),
\]
and the Toda equations (\ref{51}) give $\omega_-^\psi = 0$. The zero
curvature condition implies that
\begin{equation}
\partial_- \bW_1 = 0, \qquad \partial_- \bW_2 = 0, \label{20}
\end{equation}
and we end up with another set of characteristic integrals.

\section{Lagrangian formalism for Toda systems}

To write the action describing a Toda system we must be able to
integrate over the manifold $M$, in other words, we have to define a
volume form. To this end, as in the previous section, we treat the
manifold $M$  as a flat Riemannian manifold with a metric tensor
$\eta$. The next ingredient needed is a nondegenerate symmetric
invariant scalar product in the Lie algebra $\gothg$. We assume that
$\gothg$ is endowed with such a scalar product and denote it by $B$.

The action functional $S[\gamma]$ of a Toda system is the sum of
three terms:
\[
S[\gamma] = S_{\mathrm C}[\gamma] + S_{\mathrm{WZ}}[\gamma] +
S_{\mathrm T}[\gamma].
\]
Let us discuss them in order. 

The first term $S_{\mathrm C}[\gamma]$ is the action functional of
the principal chiral field model. Using some arbitrary coordinates 
on $M$, denoted by $z^i$, we write
\[
S_{\mathrm C}[\gamma] = - \frac{\kappa}{2} \int_M \eta^{i j}
B(\gamma^{-1} \partial_i \gamma, \gamma^{-1} \partial_j \gamma)
\sqrt{|\eta|} \, \dif^2 z,
\]
where $\kappa$ is a constant. Note that if $\gamma$ is a mapping from
$M$ to $G_0$, then $\gamma^{-1} \partial_i \gamma$ is a mapping from
$M$ to $\gothg_0$. 

The second term is the so-called Wess--Zumino term which is
constructed as follows. Suppose that the manifold $M$ is the 
boundary of the three-dimensional manifold $\widetilde M$, 
$M = \partial\widetilde M$. Let $\widetilde \gamma$ be an extension 
of the mapping $\gamma$ from $M$ to $\widetilde M$. The Wess--Zumino 
term is
\[
S_{\mathrm{WZ}}[\gamma] = - \frac{\kappa}{3!} \int_{\widetilde M}
\epsilon^{IJK} B(\widetilde \gamma^{-1} \partial_I \widetilde \gamma,
[\widetilde \gamma^{-1} \partial_J \widetilde \gamma, \widetilde
\gamma^{-1} \partial_K \widetilde \gamma]) \, \dif^3 \widetilde z,
\]
where $\widetilde z^I$ are some coordinates on $\widetilde M$ and
$\epsilon^{IJK}$ is the absolutely skew-symmetric symbol. It can
be shown that the variations of the Wess--Zumino term are determined
by the mapping $\gamma$ only. Hence the corresponding equations of
motion govern the mapping $\gamma$, leaving the extension 
$\widetilde \gamma$ arbitrary. It is an example of the so called 
multi-valued functional, and so, we write just
$S_{\mathrm{WZ}}[\gamma]$ 
instead of $S_{\mathrm{WZ}}[\widetilde \gamma]$. 

The last term is the Toda term, which has the form
\[
S_{\mathrm T}[\gamma] = \kappa \int_M B(a_-, \gamma^{-1} a_+ \gamma)
\sqrt{|\eta|} \, \dif^2 z.
\]
Here $a_-$ and $a_+$ are fixed mappings from $M$ to $\gothg_{-1}$ and
$\gothg_{+1}$, respectively, satisfying the conditions
\begin{equation}
(\sqrt{|\eta|} \, \eta^{i j} + \epsilon^{i j}) \, \partial_j a_+ = 0, 
\qquad 
(\sqrt{|\eta|} \, \eta^{i j} - \epsilon^{i j}) \,\partial_j a_- = 0. 
\label{7}
\end{equation}
The action functional of the WZNW model is the sum of the functionals
$S_{\mathrm C}[\gamma]$ and $S_{\mathrm{WZ}}[\gamma]$. The functional
$S_{\mathrm T}$ does not contain derivatives of $\gamma$. Therefore,
the construction of the Hamiltonian formalism for a Toda system is a
trivial modification of that for the WZNW model.

Let us show that the action $S[\gamma]$ does really give the Toda
equations. One finds consecutively 
\begin{eqnarray}
\delta S_{\mathrm C}[\gamma] &=& \kappa \int_M B \left(\gamma^{-1}
\delta \gamma, \frac{1}{\sqrt{|\eta|}} \, \partial_i (\sqrt{|\eta|}
\, \eta^{i j} \, \gamma^{-1} \partial_j \gamma) \right)
\sqrt{|\eta|} \, \dif^2 z, \label{79} \\
\delta S_{\mathrm{WZ}}[\gamma] &=& \kappa \int_M B
\left(\gamma^{-1} \delta \gamma, \frac{1}{\sqrt{|\eta|}} \,
\epsilon^{i j} \, \partial_i (\gamma^{-1} \partial_j \gamma)
\right) \sqrt{|\eta|} \, \dif^2 z, \label{80} \\
\delta S_{\mathrm T}[\gamma] &=& \kappa \int_M B(\gamma^{-1} \delta
\gamma, [a_-, \gamma^{-1} a_+ \gamma]) \sqrt{|\eta|} \, \dif^2 z.
\label{81}
\end{eqnarray}
To obtain from these relations the equations of motion one should use
the fact that the restriction of the scalar product $B$ to the Lie
subalgebra $\gothg_0$ is nondegenerate. To show this let us take two
elements, $x_m$ and $x_n$, belonging to $\gothg_m$ and $\gothg_n$
respectively. From (\ref{A.5}) it follows that
\[
B([x_m, q], x_n) = B(x_m, [q, x_n]),
\]
and one obtains
\[
(m + n) B(x_m, x_n) = 0.
\]
Therefore, $B(x_m, x_n) = 0$ if $n + m \ne 0$. This implies that the
restriction of the scalar product $B$ to $\gothg_0$ is nondegenerate
indeed. Note also that
\[
B|_{\gothg_{<0}} = 0, \qquad B|_{\gothg_{>0}} = 0,
\]
and that $B$ gives a nondegenerate pairing of the nilpotent 
subalgebras $\gothg_{<0}$ and $\gothg_{>0}$.

Since the scalar product $B|_{\gothg_0}$ is nondegenerate, the 
relations (\ref{79})--(\ref{81}) give rise to the following equations 
of motion
\begin{equation}
\frac{1}{\sqrt{|\eta|}} \, \partial_i (\sqrt{|\eta|} \, \eta^{i
j} \, \gamma^{-1} \partial_j \gamma + \epsilon^{i j} \, \gamma^{-1}
\partial_j \gamma) + [a_-, \gamma^{-1} a_+ \gamma] = 0. \label{8}
\end{equation}
Using light-front coordinates, one sees that these equations 
coincide with the Toda equations (\ref{1}). Here the conditions 
(\ref{7}) coincide with the conditions (\ref{6}). Rewriting the 
equations (\ref{8}) as
\[
\frac{1}{\sqrt{|\eta|}} \, \partial_i (\sqrt{|\eta|} \, \eta^{i
j} \, \partial_j \gamma \gamma^{-1} - \epsilon^{i j} \, \partial_j
\gamma \gamma^{-1}) + [\gamma a_- \gamma^{-1}, a_+] = 0, 
\]
and using light-front coordinates, we come to the equations (\ref{9}).

It is convenient now to introduce some coordinates $y^\mu$ in $G_0$ 
and work in terms of fields $\xi^\mu$ defined as
\[
\xi^\mu = y^\mu \circ \gamma = \gamma^* y^\mu.
\]
Let $g$ be the matrix-valued function which transforms the
coordinates $y^\mu(a)$ of the element $a \in G_0$ into the 
element $a$ itself, then we can write
\[
\gamma = g(\xi).
\]
Therefore, one has
\[
\gamma^{-1} \partial_i \gamma = e_\alpha \theta^\alpha_\mu(\xi)
\partial_i \xi^\mu,
\]
where $\{e_\alpha\}$ is a basis of $\gothg_0$, and the functions
$\theta^\alpha_\mu$ are defined in appendix \ref{appa}. Using this
relation, we obtain for the density of the Lagrangian of the
principal
chiral field model the expression
\[
\mathcal L_{\mathrm C} = - \frac{\kappa}{2} \sqrt{|\eta|} \,
c_{\alpha
\beta} \, \theta^\alpha_\mu(\xi) \theta^\beta_\nu(\xi) \, \eta^{i j}
\, \partial_i \xi^\mu \partial_j \xi^\nu,
\]
where the quantities $c_{\alpha \beta}$ are given by (\ref{A.9}).
Introducing the notation
\begin{equation}
h_{\mu \nu}(y) = c_{\alpha \beta} \, \theta^\alpha_\mu(y)
\theta^\beta_\nu(y), \label{24}
\end{equation}
we write the density of the Lagrangian $\mathcal L_{\mathrm C}$ as
\[
\mathcal L_{\mathrm C} = - \frac{\kappa}{2} \sqrt{|\eta|} \,
h_{\mu \nu}(\xi) \, \eta^{i j} \, \partial_i \xi^\mu \partial_j
\xi^\nu.
\]
Note that $h_{\mu \nu}(y)$ are the components of the bi-invariant
metric tensor on the Lie group~$G_0$.

The Wess--Zumino term can be written as
\[
S_{\mathrm{WZ}}[\gamma] = - \kappa \int_{\widetilde M} \widetilde
\gamma^* \Theta,
\]
where the three-form $\Theta$ is given by the relation (\ref{A.3}). 
Using the local representation (\ref{A.8}), we obtain
\[
S_{\mathrm{WZ}}[\gamma] = - \kappa \int_M \gamma^* \lambda,
\]
that gives
\[
\mathcal L_{\mathrm{WZ}} = - \frac{\kappa}{2} \, \lambda_{\mu
\nu}(\xi) \, \epsilon^{i j} \, \partial_i \xi^\mu \partial_j \xi^\nu.
\]
Finally, for the contribution to the density of the Lagrangian of the
Toda system, which is due to the term $S_{\mathrm T}[\gamma]$, we have
\[
\mathcal L_{\mathrm T} = \kappa \sqrt{|\eta|} \, B(a_-, g^{-1}(\xi)
a_+ g(\xi)) = - \kappa \sqrt{|\eta|} \, V(\xi).
\]
Collecting all terms, we come to the following expression for the
density of the Lagrangian of a Toda system
\begin{equation}
\mathcal L = - \frac{\kappa}{2} \sqrt{|\eta|} \, h_{\mu \nu}(\xi) \,
\eta^{i j} \, \partial_i \xi^\mu \partial_j \xi^\nu -
\frac{\kappa}{2}
\, \lambda_{\mu \nu}(\xi) \, \epsilon^{i j} \, \partial_i \xi^\mu
\partial_j \xi^\nu - \kappa \sqrt{|\eta|} \, V(\xi). \label{11}
\end{equation}

Let us restrict ourselves to the case when the mappings $a_-$ and
$a_+$ are constant. In this case the Toda system under consideration 
is conformally invariant, and it is possible to define the
energy-momentum tensor for it being symmetric and traceless.
Recall that there are two standard methods to define the
energy-momentum tensor. The first method is the variation of the
action over the components of the metric tensor that gives the
so-called symmetric energy-momentum tensor. Note that in our case the
Wess--Zumino term does not depend on the metric, therefore it does
not give a contribution to the symmetric energy-momentum tensor. It 
may seem to be strange at the first glance. To demonstrate that this 
is however the case, we start with the canonical energy-momentum
tensor
$T^i{}_j$ which is defined by
\[
\sqrt{|\eta|} \, T^i{}_j = - \frac{\partial \mathcal
L}{\partial(\partial_i \xi^\mu)} \, \partial_j \xi^\mu +
\delta^i_j \mathcal L.
\]
It is convenient to write the expression for the components of the
energy-momentum tensor with upper indices. We have  
\begin{eqnarray*}
\sqrt{|\eta|} \, T^{i j} &=& \kappa \sqrt{|\eta|} \, h_{\mu \nu}(\xi)
\, \eta^{i k} \eta^{j l} \partial_k \xi^\mu
\partial_l \xi^\nu - \kappa \, \lambda_{\mu \nu}(\xi) \, \epsilon^{i
k} \eta^{j l} \partial_k \xi^\mu \partial_l \xi^\nu \\
& & \hspace{1em} {} + \eta^{i j} \left(- \frac{\kappa}{2}
\sqrt{|\eta|} \, h_{\mu \nu}(\xi) \, \eta^{k l} \partial_k
\xi^\mu \partial_l \xi^\nu - \frac{\kappa}{2} \, \lambda_{\mu
\nu}(\xi) \, \epsilon^{k l} \partial_k \xi^\mu \partial_l \xi^\nu
- \kappa \sqrt{|\eta|} \, V(\xi) \right).
\end{eqnarray*}
Consider the terms containing $\lambda_{\mu \nu}(\xi)$. They can be
written as
\[
- \frac{\kappa}{2} \, (\epsilon^{i k} \eta^{l j} + \epsilon^{l i}
\eta^{k j} + \epsilon^{k l} \eta^{i j}) \lambda_{\mu \nu}(\xi) \,
\partial_k \xi^\mu \partial_l \xi^\nu.
\]
The sum $\epsilon^{i k} \eta^{l j} + \epsilon^{l i} \eta^{k j} +
\epsilon^{k l} \eta^{i j}$ is totally antisymmetric with respect to
the indices $i$, $k$ and~$l$. Since we work in a two-dimensional
space-time, this sum is equal to zero, and we can write
\[
T^{i j} = \kappa \, \eta^{i k} \eta^{j l} h_{\mu \nu}(\xi)
\, \partial_k \xi^\mu \partial_l \xi^\nu  - \frac{\kappa}{2} \,
\eta^{i j} \eta^{k l} h_{\mu \nu}(\xi) \, \partial_k \xi^\mu
\partial_l \xi^\nu - \kappa \, \eta^{\mu \nu} V(\xi).
\]
Thus, the canonical energy-momentum tensor of a Toda system has no
terms arising from the Wess--Zumino term. It can be shown that it
coincides with the symmetric energy-momentum tensor. For the
symmetric
energy-momentum tensor one has
\[
T^{i j}{}_{;i} = 0,
\]
where the usual notation for the covariant derivatives with respect
to the metric tensor $\eta$ is used. In terms of the mapping $\gamma$ 
we obtain
\[
T^{i j} = \kappa \, \eta^{i k} \eta^{j l} B(\gamma^{-1} \partial_k
\gamma, \gamma^{-1} \partial_l \gamma) - \frac{\kappa}{2} \, 
\eta^{i j} \eta^{k l} B(\gamma^{-1} \partial_k \gamma, \gamma^{-1}
\partial_l \gamma) + \kappa \, 
\eta^{i j} B(a_-, \gamma^{-1} a_+ \gamma).
\]

The trace of the obtained energy-momentum tensor is different from
zero, namely,
\[
T^i{}_i = 2 \, \kappa \, B(a_-, \gamma^{-1} a_+ \gamma).
\]
Let us construct the so-called conformally improved traceless
energy-momentum tensor. To this end first note that since the mapping
$a_-$ takes values in $\gothg_{-1}$, one can write
\[
B(a_-, \gamma^{-1} a_+ \gamma) 
= - B([q, a_-], \gamma^{-1} a_+ \gamma)
= - B(q, [a_-, \gamma^{-1} a_+ \gamma]).
\]
Taking into account the equations of motion (\ref{8}), we see that
\[
B(a_-, \gamma^{-1} a_+ \gamma) = \frac{1}{\sqrt{|\eta|}} \, 
B(q, \partial_i(\sqrt{|\eta|} \, \eta^{i j} \gamma^{-1} \partial_j
\gamma + \epsilon^{i j} \gamma^{-1} \partial_j \gamma )),
\]
that, with account of the equality
\[
[q, \gamma] = 0, 
\]
can be written as
\[
B(a_-, \gamma^{-1} a_+ \gamma) = \frac{1}{\sqrt{|\eta|}} \,
\partial_i B(q, \sqrt{|\eta|} \, \eta^{i j} \gamma^{-1}
\partial_j \gamma),
\]
Thus the trace of the energy-momentum tensor can be represented in
the
form
\[
T^i{}_i =  2 R^i{}_{;i},
\]
where
\[
R^i = \kappa B(q, \eta^{i j} \, \gamma^{-1} \partial_j \gamma) =
\kappa B(q, \eta^{i j} \, \partial_j \gamma \gamma^{-1}).
\]

Now let us use the well-known fact that the energy-momentum tensor is
defined ambiguously. In particular, one can use instead of the tensor
$T^{i j}$ the tensor
\[
T^{\prime i j} = T^{i j} + S^{i k j}{}_{;k},
\]
where the components $S^{i k j}$ satisfy the relation
\[
S^{i k j} = - S^{k i j}.
\]
It is clear that one has
\begin{equation}
T^{\prime i j}{}_{;i} = 0. \label{10}
\end{equation}
One can easily check that with the choice
\[
S^{i k j} = - 2 \eta^{i j} R^k + 2 \eta^{k j} R^i
\]
we obtain a traceless and symmetric tensor $T^{\prime i j}$. This
is the conformally improved energy-momentum tensor for the Toda 
system.

Using coordinates for which the components of the metric tensor are
constant, we come to the expression
\begin{eqnarray*}
T'_{i j} = \kappa \, B(\gamma^{-1} \partial_i \gamma,
\gamma^{-1} \partial_j \gamma) &-& \frac{\kappa}{2} \, \eta_{i j}
\eta^{k l} B(\gamma^{-1} \partial_k \gamma, \gamma^{-1} \partial_l
\gamma) \\
&+& 2 \, \kappa \, B(q, \partial_i(\gamma^{-1} \partial_j \gamma))
- \kappa \, \eta_{i j} \eta^{k l} B(q, \partial_k (\gamma^{-1}
\partial_l \gamma)).
\end{eqnarray*}
Since, the natural coordinates for a two-dimensional conformally
invariant system are light-front coordinates let us write the
components of the conformally improved energy-momentum tensor using
such coordinates. First of all recall that since the conformally
improved energy-momentum tensor is symmetric and traceless then
\[
T'_{-+} = 0, \qquad T'_{+-} = 0. 
\]
Therefore the relations (\ref{10}) take the form
\begin{equation}
\partial_+ T'_{--} = 0, \qquad \partial_- T'_{++} = 0. \label{18}
\end{equation}
It is convenient to choose the following explicit expressions for the
nonzero components:
\begin{eqnarray}
T'_{--} &=& \kappa \, B(\gamma^{-1} \partial_- \gamma, \gamma^{-1}
\partial_- \gamma) + 2 \kappa \, B(q, \partial_-(\gamma^{-1}
\partial_- \gamma)), \label{60} \\[.5em]
T'_{++} &=& \kappa \, B(\partial_+ \gamma \gamma^{-1}, \partial_+
\gamma \gamma^{-1}) + 2 \kappa \, B(q, \partial_+( \partial_+ \gamma
\gamma^{-1})). \label{61}
\end{eqnarray}

For the Toda system discussed in section \ref{te} we define
the scalar product $B$ by the relation
\begin{equation}
B(x, y) = \tr(xy). \label{33}
\end{equation}
It is clear that this scalar product is symmetric, nondegenerate
and $\Ad$-invariant. Taking into account the relations (\ref{13}),
(\ref{14}) and (\ref{15}), we obtain
\begin{eqnarray*}
T'_{--} &=& \kappa \tr \left[ \Sigma^{(1)2}_- + \Sigma^{(2)2}_- +
\partial_- (\Sigma^{(1)}_- -
\Sigma^{(2)}_-) \right], \\[.5em]
T'_{++} &=& \kappa \tr \left[ \bSigma^{(1)2}_+ + \bSigma^{(2)2}_+ +
\partial_+ (\bSigma^{(1)}_+ - \bSigma^{(2)}_+) \right].
\end{eqnarray*}
The definitions (\ref{16}) and (\ref{17}) allow us to write
\begin{equation}
T'_{--} = \frac{1}{\kappa} \, \tr \left[ W_1^2 - 2 \, W_2 \right],
\qquad T'_{++} = \frac{1}{\kappa} \, \tr \left[ \bW_1^2 - 2 \,
\bW_2 \right]. \label{44}
\end{equation}
Here the equalities (\ref{18}) can be considered as consequences of
the relations (\ref{19}) and (\ref{20}).

\section{Hamiltonian formalism}

In this section we follow mainly the paper by P.~Bowcock~\cite{Bow89}
where the Hamiltonian formulation of the WZNW model and its gauged
version was investigated. The approach used in \cite{Bow89} is based
on usage of a local representation of the closed three-form entering
the definition of the action, as an exact form. Actually we used this
trick in the previous section to construct the density of the
Lagrangian. The validity of such a local construction is justified 
by the fact that the final Hamiltonian equations do really imply the
initial Lagrangian equations.

Consider again a general Toda system. Assume that $t = z^0$ and
$x = z^1$ are flat Minkowski coordinates on $M$. In these
coordinates one has
\[
\|\eta_{i j}\| = \left( \begin{array}{cr}
-1 & 0 \\
0 & 1
\end{array} \right).
\]
The expression for the density of the Lagrangian (\ref{11}) takes the
form
\[
\mathcal L = \frac{\kappa}{2} \, h_{\mu \nu}(\xi) \, \partial_t
\xi^\mu \partial_t \xi^\nu - \frac{\kappa}{2} \, h_{\mu \nu}(\xi) \,
\partial_x \xi^\mu \partial_x \xi^\nu - \kappa \, \lambda_{\mu
\nu}(\xi) \, \partial_t \xi^\mu \partial_x \xi^\nu - \kappa V(\xi).
\]
Here and below we denote $\partial_t = \partial/\partial t$ and
$\partial_x = \partial/\partial x$. The density of the energy
functional is
\begin{equation}
\mathcal E = \frac{\partial \mathcal L}{\partial (\partial_t
\xi^\mu)}
\, \partial_t \xi^\mu - \mathcal L = \frac{\kappa}{2} \, h_{\mu
\nu}(\xi) \, \partial_t \xi^\mu \partial_t \xi^\nu + \frac{\kappa}{2}
\, h_{\mu \nu}(\xi) \, \partial_x \xi^\mu \partial_x \xi^\nu + \kappa
V(\xi). \label{12}
\end{equation}
For the generalized momenta one has the expression
\[
\pi_\mu = \frac{\partial \mathcal L}{\partial (\partial_t \xi^\mu)}
= \kappa \, h_{\mu \nu}(\xi) \, \partial_t \xi^\nu - \kappa \,
\lambda_{\mu \nu} (\xi) \, \partial_x \xi^\nu.
\]
We can write the inverse relation which expresses the generalized
velocities $\partial_t \xi^\mu$ via the generalized momenta:
\[
\partial_t \xi^\mu = \frac{1}{\kappa} \, h^{\mu \nu}(\xi) [\pi_\nu +
\kappa \, \lambda_{\nu \rho}(\xi) \, \partial_x \xi^\rho],
\]
where
\[
h^{\mu \rho}(y) \, h_{\rho \nu}(y) = \delta^\mu_\nu.
\]
Substituting the above expression for $\partial_t \xi^\mu$ into the 
relation (\ref{12}), we obtain for the density of the Hamiltonian the
following expression:
\[
\mathcal H = \frac{1}{2 \kappa} \, h^{\mu \nu}(\xi) \, [\pi_\mu +
\kappa \, \lambda_{\mu \rho}(\xi) \, \partial_x \xi^\rho] \, [\pi_\nu
+ \kappa \, \lambda_{\nu \sigma}(\xi) \, \partial_x \xi^\sigma] +
\frac{\kappa}{2} \, h_{\mu \nu}(\xi) \, \partial_x \xi^\mu \partial_x
\xi^\nu + \kappa V(\xi).
\]
Recall that the nonvanishing Poisson brackets for the fields
$\xi^\mu$ and the generalized momenta $\pi_\mu$ have the form
\[
\{\xi^\mu(x), \pi_\nu(x')\} = \delta^\mu_\nu \, \delta(x - x').
\]
Using this relation, one can write the Hamiltonian equations of
motion and prove that they are equivalent to the Lagrangian 
equations of motion.

The phase space of the system is described by the fields $\xi^\mu$
and the generalized momenta $\pi_\mu$. They depend on the choice of 
the coordinates $y^\mu$ in $G_0$. To describe the phase space in 
terms independent of this choice, consider first the quantities
\[
j_\alpha = - X_\alpha^\mu(\xi) \, [\pi_\mu + \kappa \,
\lambda_{\mu \rho}(\xi) \, \partial_x \xi^\rho] 
+ \kappa \, c_{\alpha\gamma} \, \theta^\gamma_\rho (\xi) \, 
\partial_x \xi^\rho,
\]
where the functions $X_\alpha^\mu(y)$ are defined by (\ref{A.15}). 
As is shown in appendix \ref{la}, the Poisson brackets for the
quantities $j_\alpha(x)$ are
\begin{equation}
\{j_\alpha(x), j_\beta(x')\} 
= j_\gamma(x) \, f^\gamma{}_{\alpha\beta} \, \delta(x - x') 
- 2 \kappa \, c_{\alpha \beta} \, \delta'(x - x'). \label{21}
\end{equation}
Thus, we have a realization of the so-called current algebra.

It is also convenient to consider the quantities
\[
\bar \jmath_\alpha = - \bar X_\alpha^\mu(\xi) \, [\pi_\mu +
\kappa \, \lambda_{\mu \rho}(\xi) \, \partial_x \xi^\rho] - \kappa \,
c_{\alpha \gamma} \, \bar \theta^\gamma_\rho (\xi) \, \partial_x
\xi^\rho,
\]
where the functions $\bar \theta^\alpha_\mu(y)$ are the components of
the right invariant Maurer--Cartan form of $G_0$ and the functions
$\bar X^\mu_\alpha(y)$ are defined by the equality (\ref{A.16}). 
One can show that
\begin{equation}
\{\bar \jmath_\alpha(x), \bar \jmath_\beta(x')\} = - \bar
\jmath_\gamma(x) \, f^\gamma{}_{\alpha \beta} \, \delta(x - x') 
+ 2 \kappa \, c_{\alpha \beta} \, \delta'(x - x') \label{22}
\end{equation}
and that
\begin{equation}
\{\jmath_\alpha(x), \bar \jmath_\beta(x')\} = 0. \label{23}
\end{equation}
The main relation used here is the equality (\ref{A.35}).

Now, using the definition (\ref{24}) of $h_{\mu \nu}(y)$, we obtain 
\[
h^{\mu \nu}(y) = X^\mu_\alpha(y) \, c^{\alpha \beta} \,
X^\nu_\beta(y),
\]
where
\[
c^{\alpha \gamma} c_{\gamma \beta} = \delta^\alpha_\beta.
\]
The above equality allows us to demonstrate that
\begin{eqnarray*}
c^{\alpha \beta} \, j_\alpha j_\beta = h_{\mu \nu}(\xi) [\pi_\mu &+&
\kappa \, \lambda_{\mu \rho}(\xi) \, \partial_x \xi^\rho][\pi_\nu +
\kappa \, \lambda_{\nu \sigma}(\xi) \, \partial_x \xi^\sigma] \\
&-& 2 \kappa \, \partial_x \xi^\mu [\pi_\mu + \kappa \, \lambda_{\mu
\rho}(\xi) \, \partial_x \xi^\rho] + \kappa^2 \, h_{\mu \nu}(\xi) \,
\partial_x \xi^\mu \partial_x \xi^\nu.
\end{eqnarray*}
Further, the relation (\ref{A.24}) leads to another representation of
$h_{\mu \nu}(y)$ and $h^{\mu \nu}(y)$:
\[
h_{\mu \nu}(y) = \bar \theta^\alpha_\mu(y) \, c_{\alpha \beta} \,
\bar
\theta^\beta_\nu(y), \qquad h^{\mu \nu}(y) = \bar X^\mu_\alpha(y) \,
c^{\alpha \beta} \, \bar X^\nu_\beta(y).
\]
Using these relations we find 
\begin{eqnarray*}
c^{\alpha \beta} \, \bar \jmath_\alpha \bar \jmath_\beta = h_{\mu
\nu}(\xi) [\pi_\mu &+& \kappa \, \lambda_{\mu \rho}(\xi) \,
\partial_x \xi^\rho][\pi_\nu + \kappa \, \lambda_{\nu \sigma}(\xi) \,
\partial_x \xi^\sigma] \\
&+& 2 \kappa \, \partial_x \xi^\mu [\pi_\mu + \kappa \, \lambda_{\mu
\rho}(\xi) \, \partial_x \xi^\rho] + \kappa^2 \, h_{\mu \nu}(\xi) \,
\partial_x \xi^\mu \partial_x \xi^\nu.
\end{eqnarray*}
It becomes clear that the density of the Hamiltonian $\mathcal H$ can
be written in the Sugawara form \cite{Sug68,Som68}:
\begin{equation}
\mathcal H = \frac{1}{4 \kappa} [c^{\alpha \beta} \, j_\alpha 
j_\beta + c^{\alpha \beta} \, \bar \jmath_\alpha  \bar \jmath_\beta]
+ \kappa V(\xi). \label{25}
\end{equation}

The quantities $j_\alpha$ and $\bar \jmath_\alpha$ do not depend 
on the choice of coordinates $y^\mu$ in the Lie group $G_0$ but
they depend on the choice of the basis $\{e_\alpha\}$. To get rid of
this dependence introduce the matrix-valued quantities
\[
j = e_\alpha c^{\alpha \beta} j_\beta, \qquad \bar \jmath = e_\alpha
c^{\alpha \beta} \bar \jmath_\beta.
\]
Note that $j$ and $\bar \jmath$ are the Hamiltonian counterparts of
the quantities $-\kappa \, \gamma^{-1} \partial_- \gamma$ and 
$-\kappa \, \partial_+ \gamma \gamma^{-1}$ respectively.

Our next task is to rewrite the relations (\ref{21})--(\ref{23}) in 
terms of Poisson brackets of the matrix-valued quantities $j$ and 
$\bar \jmath$. Actually we will consider $j$ and $\bar \jmath$ as
functionals on the phase space of the system taking values in the
associative algebra Mat$_N(\bbR)$. The corresponding definition of
the Poisson bracket for algebra-valued functionals on a phase space 
and its main properties are discussed in appendix \ref{av}.

Consider the element $C \in \gothg_0 \otimes \gothg_0$ defined as
\begin{equation}
C = e_\alpha \otimes e_\beta \, c^{\alpha \beta}. \label{32}
\end{equation}
Introducing the notation
\[
e^\alpha = e_\beta \, c^{\beta \alpha},
\]
we can write
\[
C = e^\alpha \otimes e^\beta \, c_{\alpha \beta} = e^\alpha
\otimes e_\alpha = e_\alpha \otimes e^\alpha.
\]
Using the relation
\[
[C, e^\gamma \otimes I_N] = [I_N \otimes e^\gamma, C] =
f^\gamma{}_{\alpha \beta} \, e^\alpha \otimes e^\beta,
\]
we obtain
\begin{eqnarray}
& \{j(x) \cmotms j(x') \} = {} - [C, I_N \otimes j(x)] \, 
\delta(x - x') - 2 \kappa \, C \, \delta'(x - x'), 
\label{27} \\[.5em]
& \{\bar \jmath(x) \cmotms \bar \jmath(x') \} = [C, I_N \otimes \bar
\jmath(x)] \, \delta(x - x') + 2 \kappa \, C \, \delta'(x - x'), 
\label{28} \\[.5em]
& \{j(x) \cmotms \bar \jmath(x') \} = 0. \label{29}
\end{eqnarray}
Using the relation (\ref{A.18}), we come to the equality
\[
\{\gamma(x), j_\alpha(x')\} = {} - \gamma(x) \, e_\alpha \,
\delta(x - x') 
\]
that gives
\begin{equation}
\{\gamma(x) \cmotms j(x')\} = {} - ( \gamma(x) \otimes I_N) \, 
C \, \delta(x - x'). \label{30}
\end{equation}
Similarly, the relation (\ref{A.19}) implies
\begin{equation}
\{\gamma(x) \cmotms \bar \jmath(x')\} 
= {} - C \, ( \gamma(x) \otimes I_N) \, \delta(x - x'). 
\label{31}
\end{equation}
It is also clear that
\[
\{ \gamma(x) \cmotms \gamma(x') \} = 0.
\]

Taking into account (\ref{25}), we obtain
\[
\mathcal H = \frac{1}{4 \kappa}
[B(j, j) + B(\bar \jmath, \bar \jmath)] 
- \kappa B(a_-, \gamma^{-1} a_+ \gamma).
\]
It is not difficult to write down the corresponding Hamiltonian
equations. If we choose as the basis quantities describing the phase
space of the system the quantities $\gamma$ and $j$, we come to the
equations
\begin{equation}
\partial_t \gamma = \partial_x \gamma - \frac{1}{\kappa} \, \gamma j,
\qquad 
\partial_t j = - \partial_x j - \kappa [a_-, \gamma^{-1} a_+ \gamma]. 
\label{34}
\end{equation}
In the case when the quantities $\gamma$ and $\bar \jmath$ are chosen
as the basis quantities, one obtains
\begin{equation}
\partial_t \gamma = - \partial_x \gamma 
- \frac{1}{\kappa} \, \bar \jmath \, \gamma, 
\qquad
\partial_t \bar \jmath = \partial_x \bar \jmath 
- \kappa [\gamma a_- \gamma^{-1}, a_+]. \label{35}
\end{equation}
It is clear that the obtained Hamiltonian equations are equivalent 
to the Toda equations (\ref{1}) and (\ref{9}) respectively.

\section{\mathversion{bold}$W$-algebra}

In this section we return again to the Toda system defined in section
\ref{te} and find the Poisson brackets for the characteristic
integrals given in section \ref{ci}. Recall that the Lie group $G_0$
in the case under consideration is isomorphic to the direct product
of
two copies of the Lie group GL$_n(\bbR)$, and the Lie
algebra $\gothg_0$ is isomorphic to the direct product of two copies
of the Lie algebra $\mathfrak{gl}_n(\bbR)$. 

The standard basis of the Lie algebra $\mathfrak{gl}_n(\bbR)$
consists
of the matrices $e_i{}^j$, $i, j = 1, \ldots, n$, defined as
\[
(e_i{}^j)^k{}_l = \delta^k_i \, \delta^j_l.
\]
Certainly, these matrices form a basis of the algebra $\Mat_n(\bbR)$, 
too. The main property of these matrices is provided by the relation
\[
e_i{}^j e_k{}^l = e_i{}^l \, \delta^j_k.
\]
Using this relation and the equality
\[
\tr (e_i^j) = \delta_i^j,
\]
one obtains
\[
\tr (e_i{}^j \, e_k{}^l) = \delta_i^l \, \delta_k^j.
\]

A natural basis of the Lie algebra $\gothg_0$ is formed by the
matrices
\[
E^{(1)}{}_i{}^j = \left( \begin{array}{cc}
e_i{}^j & 0 \\
0 & 0
\end{array} \right), \qquad
E^{(2)}{}_i{}^j = \left( \begin{array}{cc}
0 & 0 \\
0 & e_i{}^j
\end{array} \right), \qquad i,j = 1, \ldots, n.
\]
Recall that we assumed the Lie algebra $\gothg_0$ in the case
under consideration to be equipped with the scalar product $B$ 
defined by the relation (\ref{33}). Therefore, we have
\[
B(E^{(r)}{}_i{}^j, E^{(s)}{}_k{}^l) = \tr (E^{(r)}{}_i{}^j \,
E^{(s)}{}_k{}^l) = \delta_i^l \, \delta_k^j \, \delta^{rs}.
\]
{}From the natural block matrix structure of the space 
$\gothg_0 \otimes \gothg_0$, we see that the element $C$ 
introduced by (\ref{32}) has in our case the form
\begin{equation}
C = \left( \begin{array}{cc}
C_n & 0 \\
0 & C_n
\end{array} \right), \label{64}
\end{equation}
where the element 
$C_n \in \mathfrak{gl}_n(\bbR) \otimes \mathfrak{gl}_n(\bbR)$ 
is defined by the relation 
\[
C_n = e_i{}^j \otimes e_j{}^i.
\]
One can verify the validity of the equalities
\[
C_n(e_i{}^j \otimes e_k{}^l) = e_k{}^j \otimes e_i{}^l, \qquad
(e_i{}^j \otimes e_k{}^l)C_n = e_i{}^l \otimes e_k{}^j.
\]
These imply that the action of the permutation operator $P$ on
$\mathfrak{gl}_n(\bbR) \otimes \mathfrak{gl}_n(\bbR)$, or on
$\Mat_n(\bbR) \otimes \Mat_n(\bbR)$, can be realized with the help of
the element $C$ as
\[
P(a \otimes b) = C_n(a \otimes b)C_n.
\]
Note also that
\[
C_n^2 = I_n \otimes I_n.
\]

It is quite natural to use for the mapping $\gamma$ the 
parametrization (\ref{26}) and for the quantities $j$ and 
$\bar \jmath$ the parametrizations
\[
j = \left( \begin{array}{cc}
\mathcal J^{(1)} & 0 \\
0 & \mathcal J^{(2)}
\end{array} \right), \qquad 
\bar \jmath = \left( \begin{array}{cc}
\bar {\mathcal J}^{(1)} & 0 \\
0 & \bar {\mathcal J}^{(2)}
\end{array} \right),
\]
where the functions $\mathcal J^{(r)}$ and $\bar {\mathcal J}^{(r)}$,
$r = 1,2$, take values in $\mathfrak{gl}_n(\bbR)$. It is clear that
the relations (\ref{27})--(\ref{29}) can be written now as
\begin{eqnarray}
& \{\mathcal J^{(r)}(x) \cmotms \mathcal J^{(s)}(x') \} = {} - \left(
[C_n, I_n \otimes \mathcal J^{(r)}(x)] \, \delta(x - x') 
+ 2 \kappa \, C_n \, \delta'(x - x') \right) \delta^{rs}, 
\label{37} \\[.5em]
& \{\bar \mathcal J^{\,(r)}(x) \cmotms \bar \mathcal J^{\,(s)}(x') \}
= \left([C_n, I_n \otimes \bar \mathcal J^{\, (r)}(x)] \,
\delta(x - x') + 2 \kappa \, C_n \, \delta'(x - x') \right)
\delta^{rs}, \label{38} \\[.5em]
& \{\mathcal J^{(r)}(x) \cmotms \bar \mathcal J^{\, (s)}(x') \} = 0,
\label{39}
\end{eqnarray}
and the relations (\ref{30}) and (\ref{31}) take the forms 
\begin{eqnarray}
\{\Gamma^{(r)}(x) \cmotms \mathcal J^{(s)}(x')\} &=& {} - (
\Gamma^{(r)}(x) \otimes I_n) \, C_n \, \delta(x - x') \, \delta^{rs},
\label{40} \\
\{\Gamma^{(r)}(x) \cmotms \bar \mathcal J^{(s)}(x')\} &=& {} - C_n \,
(\Gamma^{(r)}(x) \otimes I_n) \, \delta(x - x') \, \delta^{rs}.
\label{41}
\end{eqnarray}
The Hamiltonian equations of motion (\ref{34}) are now of the forms 
\begin{eqnarray*}
& \partial_t \Gamma^{(1)} = \partial_x \Gamma^{(1)} -
\frac{1}{\kappa} \, \Gamma^{(1)} \mathcal J^{(1)}, \qquad \partial_t
\mathcal J^{(1)} = - \partial_x \mathcal J^{(1)} + \kappa \,
\Gamma^{(1)-1} \Gamma^{(2)}, \\
&\partial_t \Gamma^{(2)} = \partial_x \Gamma^{(2)} - \frac{1}{\kappa}
\, \Gamma^{(2)} \mathcal J^{(2)}, \qquad \partial_t \mathcal J^{(2)}
= - \partial_x \mathcal J^{(2)} - \kappa \, \Gamma^{(1)-1}
\Gamma^{(2)},
\end{eqnarray*}
while for the equations (\ref{35}) we have
\begin{eqnarray*}
& \partial_t \Gamma^{(1)} = - \partial_x \Gamma^{(1)} -
\frac{1}{\kappa} \, \bar \mathcal J^{(1)} \Gamma^{(1)}, \qquad
\partial_t \bar \mathcal J^{(1)} = \partial_x \bar \mathcal J^{(1)} +
\kappa \, \Gamma^{(2)} \Gamma^{(1)-1}, \\
& \partial_t \Gamma^{(2)} = - \partial_x \Gamma^{(2)} -
\frac{1}{\kappa} \, \bar \mathcal J^{(2)} \Gamma^{(2)}, \qquad
\partial_t \bar \mathcal J^{(2)} = \partial_x \bar \mathcal J^{(2)} -
\kappa \, \Gamma^{(2)} \Gamma^{(1)-1}.
\end{eqnarray*}

Let us find Hamiltonian counterparts of the characteristic
integrals $W_1$ and $W_2$ defined by the relation (\ref{16}). 
There is no problem with the characteristic integral $W_1$. 
Its Hamiltonian counterpart obviously is
\[
\cW_1 = \mathcal J^{(1)} + \mathcal J^{(2)}.
\]
The characteristic integral $W_2$ contains higher time derivatives
and
has no direct Hamiltonian counterpart. However, here one can use 
the fact that characteristic integrals are defined up to terms
vanishing at the equations of motion. Therefore, one can use
the equations of motion to get equivalent characteristic integrals
which do not contain higher time derivatives.

For the case under consideration, using the definition (\ref{14}),
the equations of motion (\ref{36}) and the equality
\[
\partial_- = \partial_+ - 2 \, \partial_x,
\]
we obtain
\[
\frac{1}{2} \, (\partial_- \Sigma^{(1)}_- - \partial_-
\Sigma^{(2)}_-) = - \Gamma^{(1)-1} \Gamma^{(2)} - (\partial_x
\Sigma^{(1)}_- - \partial_x \Sigma^{(2)}_-).
\]
Hence, the Hamiltonian counterpart of the characteristic integral
$W_2$ is
\[
\cW_2 = \mathcal J^{(1)} \mathcal J^{(2)} - \kappa \,
(\partial_x \mathcal J^{(1)} - \partial_x \mathcal J^{(2)}) +
\kappa^2 \, \Gamma^{(1)-1} \Gamma^{(2)}.
\]

The Poisson bracket for the characteristic integral $\cW_1$
follows directly from (\ref{37}):
\begin{equation}
\{\cW_1(x) \cmotms \cW_1(x')\} = {}- [C_n, I_n \otimes
\cW_1(x)] \, \delta (x - x') - 4 \kappa \, C_n \, 
\delta'(x - x'). \label{55}
\end{equation}
The calculations needed to obtain expressions for other Poisson
brackets are more complicated. The main formulas are presented in
appendix \ref{wc}. The final result is
\begin{eqnarray}
\{\cW_1(x) && {} \cmotms \cW_2(x') \} = {} - [C_n, I_n
\otimes \cW_2(x)] \, \delta(x - x') \nonumber \\*
&& \hspace{10em} {} - \kappa \, [C_n, I_n \otimes \cW_1(x')]_+
\, \delta'(x - x'), \label{42} \\[.3em]
\{\cW_2(x) && {} \cmotms \cW_2(x')\} = (I_n \otimes \cW_2(x)) \, C_n
\, (I_n \otimes \cW_1(x)) \, \delta(x - x')  \nonumber \\
&& {} - (I_n \otimes \cW_1(x)) \, C_n \, (I_n \otimes \cW_2(x)) \,
\delta(x - x') \nonumber \\[.3em]
&& {} - \frac{\kappa^2}{2} \, [C_n, I_n \otimes \partial_x^2 \mathcal
W_1(x)] \, \delta(x - x') \nonumber \\[.3em]
&& {} + \kappa \, [C_n, I_n \otimes (\cW_2(x) 
+ \mathcal W_2(x'))]_+ \, \delta'(x - x') \nonumber \\[.5em]
&& {} - \kappa \, (I_n \otimes \cW_1(x)) \, C_n \, (I_n \otimes
\cW_1(x)) \, \delta'(x - x') \nonumber \\[.3em]
&& {} - \kappa \, (I_n \otimes \cW_1(x')) \, C_n \, (I_n \otimes
\cW_1(x')) \, \delta'(x - x') \nonumber \\[.3em]
&& {} + \frac{3 \kappa^2}{2} \, 
[C_n, I_n \otimes (\cW_1(x) + \cW_1(x'))] \, \delta''(x - x') 
+ 4 \kappa^3 \, C_n \, \delta'''(x - x'). \label{43}
\end{eqnarray}
Some terms of the last formula can be combined into a commutator.
Actually it would give a more compact expression. Nevertheless, we
prefer to use the above form of the expression which is more
convenient for the comparison with the case considered in the next
section. After some redefinitions one can get convinced that
the obtained expressions for the Poisson brackets of the
characteristic integrals coincide with the expressions obtained via
the method of Hamiltonian reduction \cite{BTV91}. Our direct
rederivation of these expressions can be considered, in particular, as
the verification needed by the reasons given in the introduction.

The Hamiltonian counterparts of the characteristic integrals $\bW_1$
and $\bW_2$ are
\begin{eqnarray*}
& \bcW_1 = \bcJ{(1)} + \bcJ{(2)}, \\[.3em]
& \bcW_2 = \bcJ{(2)} \bcJ{(1)} + \kappa \, (\partial_x \bcJ{(1)} -
\partial_x \bcJ{(2)}) + \kappa^2 \Gamma^{(2)} \Gamma^{(1)-1},
\end{eqnarray*}
and the Poisson brackets for them look as
\begin{eqnarray}
\{ \bcW_1(x) && \cmotms \bcW_1(x') \} = [C_n, I_n \otimes \bcW_1(x)]
\, \delta (x - x') + 4 \kappa \, C_n \, \delta'(x - x'), \label{52}
\\
\{\bcW_1(x) && {} \cmotms \bcW_2(x') \} = [C_n, I_n \otimes
\bcW_2(x)] \, \delta(x - x') \nonumber \\*
&& \hspace{10em} {} + \kappa \, [C_n, I_n \otimes \bcW_1(x')]_+
\, \delta'(x - x'), \label{53} \\[.3em]
\{\bcW_2(x) && {} \cmotms \bcW_2(x')\} = {}- (I_n \otimes \bcW_2(x))
\, C_n \, (I_n \otimes \bcW_1(x)) \, \delta(x - x') \nonumber \\
&& {} + (I_n \otimes \bcW_1(x)) \, C_n \, (I_n \otimes \bcW_2(x)) \,
\delta(x - x') \nonumber \\[.3em]
&& {} + \frac{\kappa^2}{2} \, [C_n, I_n \otimes \partial_x^2 
\bcW_1(x)] \, \delta(x - x') \nonumber \\[.3em]
&& {} - \kappa \, [C_n, I_n \otimes (\bcW_2(x) + \bcW_2(x'))]_+ \,
\delta'(x - x') \nonumber \\[.5em]
&& {} + \kappa \, (I_n \otimes \bcW_1(x)) \, C_n \, (I_n \otimes
\bcW_1(x)) \, \delta'(x - x') \nonumber \\[.3em]
&& {} + \kappa \, (I_n \otimes \bcW_1(x')) \, C_n \, (I_n \otimes
\bcW_1(x')) \, \delta'(x - x') \nonumber \\[.3em]
&& {} - \frac{3 \kappa^2}{2} \, 
[C_n, I_n \otimes (\bcW_1(x) + \bcW_1(x'))] \, \delta''(x - x') 
- 4 \kappa^3 \, C_n \, \delta'''(x - x'). \label{54}
\end{eqnarray}
Here we again write the result of our calculations in the form which
is convenient from the point of view of the example considered in the
next section.

Let us find the Poisson bracket for the Hamiltonian counterparts
$\mathcal T'_{--}$ and $\mathcal T'_{++}$ of the components $T'_{--}$
and $T'_{++}$ of the energy-momentum tensor. It is known that they
are
the generators of the conformal transformations. 

As follows from (\ref{44}), one has
\[
\mathcal T'_{--} = \frac{1}{\kappa} \, \tr \left[ \cW_1^2 - 2 \cW_2
\right], \qquad \mathcal T'_{++} = \frac{1}{\kappa} \, \tr \left[
\bcW_1^2 - 2 \bcW_2 \right].
\]
To find the Poisson brackets in question, we start with the relation
\begin{equation}
\{ \cW_1^2(x) \cmotms \cW_1(x') \} = {} - [C_n, I_n \otimes
\cW_1^2(x)] \,
\delta(x - x') - 4 \kappa \, [C_n, I_n \otimes \cW_1(x)]_+ \,
\delta'(x - x'). \label{49}
\end{equation}
This relation gives
\begin{eqnarray}
\{ \cW_1^2(x) && \cmotms \cW_1^2(x') \} = {} - [C_n, I_n \otimes
\cW_1^3(x)] \, \delta(x - x') \nonumber \\*
&& {} - [C_n, \cW_1(x) \otimes \cW_1^2(x)] \, \delta(x - x')
\nonumber \\*[.5em]
&& {} - 2 \kappa \, [C_n, I_n \otimes (\cW_1(x) \partial_x \cW_1(x) -
\partial_x \cW_1(x) \cW_1(x))] \, \delta(x - x') \nonumber \\*[.5em]
&& {} - 2 \kappa \, [C_n, I_n \otimes (\cW_1^2(x) + \cW_1^2(x'))]_+
\, \delta'(x - x') \nonumber \\*[.5em]
&& {} - 4 \kappa \, C_n \, (\cW_1(x) \otimes \cW_1(x) + \cW_1(x')
\otimes \cW_1(x')) \, \delta'(x - x'). \label{45}
\end{eqnarray}
For any $\Mat_n(\bbR)$-valued functionals $F$ and $G$ one obtains
\[
\{ \tr F, \tr G \} = \tr \{F \cmotms G \}.
\]
Besides, for any $a, b \in \Mat_n(\bbR)$ one has
\[
\tr (a \otimes b) = \tr a \, \tr b,
\]
and
\[
\tr (C_n (a \otimes b)) = \tr ((a \otimes b) C_n) = \tr (ab).
\]
Using these relations, we obtain from (\ref{45}) the equality
\begin{equation}
\{ \tr \cW_1^2(x), \tr \cW_1^2(x') \} = {} - 8 \kappa \, (\tr
\cW_1^2(x) + \tr \cW_1^2(x')) \, \delta'(x - x'). \label{46}
\end{equation}
The relation
\begin{eqnarray}
\{ \cW_1^2(x) && \cmotms \cW_2(x') \} = {} - [C_n, \cW_1(x) \otimes
\cW_2(x)] \, \delta(x - x') \nonumber \\
&& {} - C_n \, (I_n \otimes \cW_1(x) \cW_2(x) - \cW_2(x) \cW_1(x)
\otimes I_n ) \, \delta(x - x') \nonumber \\[.5em]
&& - \kappa \, [C_n, \cW_1(x) \otimes \cW_1(x')]_+ \, \delta'(x - x')
\nonumber \\[.5em]
&& - \kappa \, C_n \, (I_n \otimes \cW_1(x) \cW_1(x') 
+ \cW_1(x') \cW_1(x) \otimes I_n) \, \delta'(x - x') 
\label{50}
\end{eqnarray}
helps us to obtain that
\begin{equation}
\{ \tr \cW_1^2(x), \tr \cW_2(x') \} = {} - 2 \kappa \, (\tr \cW_1^2(x)
+ \tr \cW_1^2(x')) \, \delta'(x - x'). \label{47}
\end{equation}
Further, the relation (\ref{43}) gives
\begin{eqnarray}
\{ \tr \cW_2(x), {} && \tr \cW_2(x') \} = 2 \kappa \, 
(\tr \cW_2(x) + \tr \cW_2(x')) \, \delta'(x - x') 
\nonumber \\[.3em]
&& {} - \kappa \, (\tr \cW_1^2(x) + \tr \cW_1^2(x')) \,
\delta'(x - x') + 4 \kappa^3 \, n \, \delta'''(x - x'). \label{48}
\end{eqnarray}
Taking into account the relations (\ref{46}), (\ref{47}) and
(\ref{48})
we get 
\[
\{ \mathcal T'_{--}(x), \mathcal T'_{--}(x') \} = {} - 4 \, 
(\mathcal T'_{--}(x) + \mathcal T'_{--}(x')) \, \delta'(x - x') 
+ 16 \kappa \, n \, \delta'''(x - x').
\]
In a similar way we come to the equality
\[
\{ \mathcal T'_{++}(x), \mathcal T'_{++}(x') \} = 4 \, 
(\mathcal T'_{++}(x) + \mathcal T'_{++}(x')) \, \delta'(x - x') 
- 16 \kappa \, n \, \delta'''(x - x'),
\]
and it is evident that
\[
\{ \mathcal T'_{--}(x), \mathcal T'_{++}(x') \} = 0.
\]
It is clear from these relations that the quantities
\begin{equation}
\mathcal V(x) = \frac{1}{4} \, \mathcal T'_{--}(x), \qquad 
\bar \mathcal V(x) = \frac{1}{4} \, \mathcal T'_{++}(x) \label{67}
\end{equation}
are generators of two copies of the Virasoro algebra:
\begin{eqnarray*}
& \{\mathcal V(x), \mathcal V(x')\} = {} - (\mathcal V(x) + 
\mathcal V(x')) \, \delta'(x - x') + \kappa \, n \, \delta'''(x - x'),
\\[.3em]
& \{\bar \mathcal V(x), \bar \mathcal V(x')\} 
= (\bar \mathcal V(x) + \bar \mathcal V(x')) \, \delta'(x - x') 
- \kappa \, n \, \delta'''(x - x'), \\[.3em]
& \{ \mathcal V(x), \bar \mathcal V(x') \} = 0.
\end{eqnarray*}

The generators $\mathcal V(x)$ and $\bar \mathcal V(x)$ produce
infinitesimal conformal transformations via the following standard
procedure. Let us define
\[
\mathcal V_\varepsilon(t) = \int \dif x \, \varepsilon(t, x) \,
\mathcal V(t, x),
\]
where $\varepsilon$ is an arbitrary infinitesimal function on $M$
which satisfies the relation
\[
\partial_+ \varepsilon = \partial_t \varepsilon + \partial_x
\varepsilon = 0.
\]
Actually, $\mathcal V_\varepsilon$ is an integrated characteristic
integral, therefore, it does not depend on $t$. Consider the
infinitesimal transformations defined for an arbitrary observable
$F(t)$ as
\[
\delta F(t) = \{\mathcal V_\varepsilon(t), F(t)\}.
\]
It can be shown that these transformations are the infinitesimal
version of the conformal transformations described by the relation
(\ref{71}), (\ref{72}) with $\zeta^+(z^+) = z^+$. Similarly, the 
quantities
\[
\bar \mathcal V_{\bar \varepsilon}(t) = \int \dif x \, \bar
\varepsilon(t, x) \, \bar \mathcal V(t, x),
\]
where
\[
\partial_- \bar \varepsilon = \partial_t \bar \varepsilon -
\partial_x \bar \varepsilon = 0,
\]
generate the infinitesimal conformal transformations described by
the relation (\ref{71}), (\ref{72}) with $\zeta^-(z^-) = z^-$.

Now we will find the conformal weights of $\cW_1$ and $\cW_2$. Recall
that a field $\Phi(x)$ has the conformal weight $h$ with respect to
$\mathcal V(x)$ if 
\[
\{\mathcal V(x), \Phi(x')\} = {} - (\Phi(x) + (h - 1)\Phi(x')) \,
\delta'(x - x') + \ldots,
\]
where the dots stand for possible central terms, and it has the
conformal weight $\bar h$ with respect to $\bar \mathcal V(x)$ if
\[
\{\bar \mathcal V(x), \Phi(x')\} = (\Phi(x) + (\bar h - 1)\Phi(x'))
\, \delta'(x - x') + \ldots.
\]

Introduce two mappings $\tr_1$ and $\tr_2$ from $\Mat_n(\bbR) \otimes
\Mat_n(\bbR)$ to $\Mat_n(\bbR)$ given by
\[
\tr_1 (a \otimes b) = (\tr a) b, \qquad 
\tr_2 (a \otimes b) = a (\tr b).
\]
It can be verified that these mappings satisfy the relations
\begin{eqnarray*}
& \tr_1 (C_n (a \otimes b)) = ab, \qquad 
\tr_1 ((a \otimes b) C_n) = ba, \\[.3em]
& \tr_2 (C_n (a \otimes b)) = ba, \qquad 
\tr_2 ((a \otimes b) C_n) = ab.
\end{eqnarray*}
For any $\Mat_n(\bbR)$-valued functionals $F$ and $G$ one has
\[
\tr_1 \{F \cmotms G \} = \{ \tr F \cmotms G\}, \qquad \tr_2 \{F
\cmotms G \} = \{ F \cmotms \tr G\}.
\]
Using the above relations, one obtains from (\ref{49}) and (\ref{42})
the equalities
\begin{eqnarray*}
& \{ \tr \cW_1^2(x) \cmotms \cW_1(x') \} = {} - 8 \kappa \, \cW_1(x)
\, \delta (x - x'), \\
& \{ \tr \cW_2(x) \cmotms \cW_1(x') \} = {} - 2 \kappa \, \cW_1(x) \,
\delta (x - x').
\end{eqnarray*}
Hence, we come to the relation
\[
\{ \mathcal V(x) \cmotms \cW_1(x') \} = {} - \cW_1(x) \, 
\delta'(x - x').
\]

{}From the relation (\ref{50}) we obtain
\begin{eqnarray*}
\{\tr \cW_1^2(x) \cmotms \cW_2(x') \} = {} - 2 \, (\cW_1(x) \cW_2(x)
&-& \cW_2(x) \cW_1(x)) \, \delta(x - x') \\
&-& 2 \kappa \, (\cW_1^2(x) + \cW_1^2(x')) \, \delta'(x - x'),
\end{eqnarray*}
and the equality (\ref{43}) gives
\begin{eqnarray*}
\{ && \tr \cW_2(x) \cmotms \cW_2(x') \} = - (\cW_1(x) \cW_2(x)
- \cW_2(x) \cW_1(x)) \, \delta(x - x') \\
&& {} - \kappa \, (\cW_1^2(x) + \cW_1^2(x')) \, \delta'(x - x') 
+ 2 \kappa \, (\cW_2(x) + \cW_2(x')) \, \delta'(x - x') 
+ 4 \kappa^3 \, I_n \, \delta'''(x - x').
\end{eqnarray*}
Consequently, one has
\[
\{ \mathcal V(x) \cmotms \cW_2(x') \} = {} - (\cW_2(x) + \cW_2(x'))
\, \delta'(x - x') - 2 \kappa^2 \, \delta'''(x - x').
\]
Thus, the characteristic integral $\cW_1$ has the conformal weight 1
and the characteristic integral $\cW_2$ has the conformal weight 2
with respect to $\mathcal V(x)$. Similarly, we obtain that the
characteristic integral $\bcW_1$ has the conformal weight 1 and the
characteristic integral $\bcW_2$ has the conformal weight 2 with
respect to $\bar \mathcal V(x)$.

In the end of this section we find the form of the infinitesimal
symmetry transformations generated by the characteristic integrals.
First consider the quantity
\[ 
\mathcal W_\varepsilon(t) = \int \dif x \, \tr \left[
\varepsilon_1(t, x) \mathcal W_1 (t, x) 
+ \varepsilon_2(t, x) \mathcal W_2(t, x) \right],
\] 
where $\varepsilon_1$ and $\varepsilon_2$ are arbitrary infinitesimal
matrix-valued functions on $M$ satisfying the relations
\[
\partial_+ \varepsilon_1 = \partial_t \varepsilon_1 + \partial_x
\varepsilon_1 = 0, 
\qquad 
\partial_+ \varepsilon_2 = \partial_t \varepsilon_2 + \partial_x
\varepsilon_2 = 0.
\]
The infinitesimal transformations generated by $\cW_\varepsilon(t)$
written in the Lagrangian form are
\begin{eqnarray}
&& \delta_\varepsilon \Gamma^{(1)} = \Gamma^{(1)} \, \varepsilon_1 -
\kappa \, \Gamma^{(1)} \, \Gamma^{(2)-1} \, \partial_- \Gamma^{(2)}
\,
\varepsilon_2 - \frac{\kappa}{2} \, \Gamma^{(1)} \, \partial_-
\varepsilon_2, \label{73} \\
&& \delta_\varepsilon \Gamma^{(2)} = \Gamma^{(2)} \, \varepsilon_1 -
\kappa \, \Gamma^{(2)} \, \varepsilon_2 \, \Gamma^{(1)-1} \,
\partial_- \Gamma^{(1)} + \frac{\kappa}{2} \, \Gamma^{(2)} \,
\partial_- \varepsilon_2. \label{74}
\end{eqnarray}
According to (\ref{55})--(\ref{43}), the generators $\cW_\varepsilon$
satisfy the relation
\begin{equation}
\{ \mathcal{W}_\mu , \mathcal{W}_\nu \} {} 
= \mathcal{W}_{\varepsilon(\mu,\nu)} + \mathcal{C}(\mu,\nu), 
\label{WmWn}
\end{equation}
with the infinitesimal matrix-valued functions and the central 
extension term being   
\begin{eqnarray*}
\varepsilon_1(\mu,\nu) &=& [\mu_1 , \nu_1] + \kappa\left( [\partial_x
\mu_1 , \nu_2]_+ + [\mu_2 , \partial_x\nu_1]_+ \right) \\[.5em]
&-& \kappa \left( \partial_x \nu_2 \, \mathcal{W}_1 \, \mu_2 - \nu_2
\, \mathcal{W}_1 \, \partial_x\mu_2 \right) - \kappa^2
\left([\partial^2_x\mu_2 , \nu_2] - [\partial_x\mu_2 ,
\partial_x\nu_2] + [\mu_2 , \partial^2_x\nu_2] \right), \\[.5em]
\varepsilon_2(\mu,\nu) &=& [\mu_1 , \nu_2] + [\mu_2 , \nu_1] 
+ \left( \mu_2 \, \mathcal{W}_1 \, \nu_2 - \nu_2 \, \mathcal{W}_1 \,
\mu_2 \right) + \kappa \left( [\mu_2 , \partial_x\nu_2]_+ 
- [\partial_x\mu_2 , \nu_2]_+ \right), \\[.5em]
\mathcal{C}(\mu,\nu) &=& 4\kappa \int \dif x \, \tr \left(\partial_x
\mu_1 \, \nu_1\right) - 4\kappa^3 \int \dif x \, \tr
\left(\partial^3_x \mu_2 \, \nu_2\right).  
\end{eqnarray*}
We see that the nonlinear terms of the $W$-algebra made the
transformation parameters $\varepsilon_1$ and $\varepsilon_2$ 
depending on the Toda fields and their derivatives, although only
through the characteristic integral $\mathcal W_1$.

Similarly, introducing the quantity
\[ 
\bcW_{\bar\varepsilon}(t) = \int \dif x \, \tr \left[
\bar \varepsilon_1(t, x) \bcW_1 (t, x) 
+ \bar \varepsilon_2(t, x) \bcW_2(t, x) \right],
\]
where the infinitesimal matrix-valued functions 
$\bar \varepsilon_1$ and $\bar \varepsilon_2$ satisfy the relations
\[
\partial_- \bar \varepsilon_1 
= \partial_t \bar \varepsilon_1 - \partial_x \bar \varepsilon_1 = 0, 
\qquad 
\partial_- \bar \varepsilon_2 
= \partial_t \bar \varepsilon_2 - \partial_x \bar \varepsilon_2 = 0,
\]
we come to the following expressions for the infinitesimal
transformations
\begin{eqnarray}
&& \delta_{\bar \varepsilon} \Gamma^{(1)} = \bar \varepsilon_1 \,
\Gamma^{(1)} - \kappa \, \bar \varepsilon_2 \, \partial_+
\Gamma^{(2)}
\, \Gamma^{(2)-1} \, \Gamma^{(1)}  - \frac{\kappa}{2} \, \partial_+
\bar \varepsilon_2 \, \Gamma^{(1)}, \label{75} \\
&& \delta_{\bar \varepsilon} \Gamma^{(2)} = \bar \varepsilon_1 \,
\Gamma^{(2)} - \kappa \, \partial_+ \Gamma^{(1)} \, \Gamma^{(1)-1} \,
\bar \varepsilon_2 \, \Gamma^{(2)} + \frac{\kappa}{2} \, \partial_+
\bar \varepsilon_2 \, \Gamma^{(2)}. \label{76}
\end{eqnarray}
The generators $\bcW_{\bar \varepsilon}$ give rise to a closed algebra
of the form (\ref{WmWn}), that can be found from the relations
(\ref{52})--(\ref{54}). Actually, we have 
\begin{equation}
\{ \bcW_{\bar \mu} , \bcW_{\bar \nu} \} {} 
= \bcW_{\bar \varepsilon(\bar\mu,\bar\nu)} +
\bar\mathcal{C}(\bar\mu,\bar\nu), 
\label{bWmbWn}
\end{equation}
where 
\begin{eqnarray*}
\bar \varepsilon_1(\bar \mu, \bar \nu) &=& - [\bar\mu_1 , \bar\nu_1] 
- \kappa\left( [\partial_x \bar\mu_1 , \bar\nu_2]_+ + [\bar\mu_2 ,
\partial_x \bar\nu_1]_+ \right) \\[.5em]
&+& \kappa\left( \partial_x \bar\nu_2 \, \bcW_1 \, \bar\mu_2 -
\bar\nu_2 \, \bcW_1 \, \partial_x \bar\mu_2 \right) + \kappa^2\left(
[\partial^2_x \bar\mu_2 , \bar\nu_2]  - [\partial_x\bar\mu_2 ,
\partial_x \bar\nu_2] + [\bar\mu_2 , \partial^2_x \bar\nu_2] \right),
\\[.5em]
\bar\varepsilon_2(\bar\mu,\bar\nu) &=& - [\bar\mu_1 , \bar\nu_2] 
- [\bar\mu_2 , \bar\nu_1] - \left( \bar\mu_2 \, \bcW_1 \, \bar\nu_2 -
\bar\nu_2 \, \bcW_1 \, \bar\mu_2 \right) - \kappa\left( [\bar\mu_2 ,
\partial_x \bar\nu_2]_+ - [\partial_x\bar\mu_2 , \bar\nu_2]_+ \right),
\\[.5em]
\bar\mathcal{C}(\bar\mu,\bar\nu) &=& - 4\kappa \int \dif x \, 
\tr \left(\partial_x \bar\mu_1 \, \bar\nu_1\right) + 4\kappa^3 \int
\dif x \, \tr \left(\partial^3_x \bar\mu_2 \, \bar\nu_2\right).  
\end{eqnarray*} 
One can verify that the transformations (\ref{73}), (\ref{74}) 
and (\ref{75}), (\ref{76}) are symmetry transformations for the
Toda system under consideration. Putting $\varepsilon_2 = 0$ and
$\bar\varepsilon_2 = 0$ we obtain the infinitesimal version of 
the transformations described by the relation (\ref{77}).

\section{Non-abelian Liouville equation} \label{nle}

In this section we consider an example of a non-abelian Toda system
associated with the Lie group Sp$_{n}(\bbR)$. It is convenient for
our purposes to define this Lie group as a subgroup of the Lie group
GL$_{\,2n}(\bbR)$ formed by all matrices $a \in \mathrm{GL}_{\,
2n}(\bbR)$ which satisfy the relation
\[
a^t \, K_n \, a = K_n,
\]
where the $2n \times 2n$ matrix $K_n$ has the form
\[
K_n = \left( \begin{array}{cc}
0 & J_n \\
-J_n & 0
\end{array} \right)
\]
with $J_n$ being the skew-diagonal unit $n \times n$ matrix. The
superscript $t$ as usually means the transposition.

The Lie algebra $\mathfrak{sp}_n(\bbR)$ of the Lie group
$\mathrm{Sp}_n(\bbR)$ is formed by all real $2n \times 2n$ matrices
$x$ which satisfy the relation
\[
x^t K_n + K_n \, x = 0.
\]
Using for a general $2n \times 2n$ matrix $x$ block representation
(\ref{56}), we see that the above relation is equivalent to the
equalities
\[
x_{11}^T = -x_{22}^{}, \qquad x_{12}^T = x_{12}^{}, \qquad x_{21}^T =
x_{21}^{},
\]
where for an $n \times n$ matrix $x$ we have denoted 
\[
x^T = (J_n)^{-1} \, x^t J_n = J_n \, x^t J_n.
\]
Actually the matrix $x^T$ is the transpose of the matrix $x$ with
respect to the main skew diagonal.

Note that the matrix $q$ defined by (\ref{13}) belongs to
$\mathfrak{sp}_n(\bbR)$ and defines its $\bbZ$-gradation. Consider
the
Toda system associated with this $\bbZ$-gradation.

One can verify that the Lie group $G_0$ in the case under
consideration is formed by the block $2n \times 2n$ matrices $a$ of
the form
\[
a = \left( \begin{array}{cc}
b & 0 \\
0 & (b^T)^{-1}
\end{array} \right),
\]
where $b$ is an arbitrary element of $\mathrm{GL}_n(\bbR)$. 
Hence, the subgroup $G_0$ is isomorphic to the Lie group
$\mathrm{GL}_n(\bbR)$, and the mapping $\gamma$ entering 
the general Toda equations (\ref{1}), can be parametrized 
as
\begin{equation}
\gamma = \left( \begin{array}{cc}
\Gamma & 0 \\
0 & (\Gamma^T)^{-1}
\end{array} \right), \label{65}
\end{equation}
where the mapping $\Gamma$ takes values in $\mathrm{GL}_n(\bbR)$. The
general form of the mappings $a_-$ and $a_+$ is again given by
(\ref{57}). Here the mappings $A_-$ and $A_+$ must satisfy the
relations (\ref{58}) and
\[
A_-^T = A_-^{}, \qquad A_+^T = A_+^{}.
\]
Putting $A_- = A_+ = I_n$ we come to the following Toda equations
\begin{equation}
\partial_+ (\Gamma^{-1} \partial_- \Gamma) = - (\Gamma^T \Gamma)^{-1}.
\label{59}
\end{equation}
In the case of $n=1$ the mapping $\Gamma$ is just an ordinary
function on $M$ taking values in $\bbR^\times$. If the function 
$\Gamma$ is continuous, then it is either positive or negative. 
For a positive function $\Gamma$ one can write $\Gamma = \exp F$ 
and the equation (\ref{59}) takes the form
\[
\partial_+ \partial_- F = - \exp(-2 F),
\]
that is the well-known Liouville equation. Therefore, it is natural
to call the matrix differential equation (\ref{59}) the non-abelian 
Liouville equation.

The system under consideration possesses a WZNW-type symmetry
\begin{equation}
\Gamma \to \Lambda_+ \Gamma \Lambda^{}_-, \label{83}
\end{equation}
where the matrix-valued functions $\Lambda_-$ and $\Lambda_+$ satisfy
the conditions
\begin{eqnarray*}
& \partial_+ \Lambda_- = 0, \qquad \partial_- \Lambda_+ = 0, \\[.3em]
& \Lambda_-^T = \Lambda_-^{-1}, \qquad \Lambda_+^T = \Lambda_+^{-1}.
\end{eqnarray*}
It is also conformally invariant. Here the action of the conformal
transformations on $\Gamma$ is defined as
\begin{equation}
\Gamma(z^-, z^+) \to 
[\partial_- \zeta^-(z^-) \partial_+ \zeta^+(z^+)]^{-1/2}
\Gamma(\zeta^+(z^+), \zeta^-(z^-)). \label{82}
\end{equation}

The procedure described in section \ref{ci}, leads now to the
following matrix characteristic integrals
\begin{eqnarray}
& W_1 = - \frac{\kappa}{2} \, (\Sigma_- - \Sigma_-^T), \qquad W_2 = -
\kappa^2 \left( \frac{1}{2} \, \partial_-(\Sigma_- + \Sigma_-^T) +
\Sigma_- \Sigma_-^T \right), \label{62} \\
& \bW_1 = - \frac{\kappa}{2} \, (\bSigma_+ - \bSigma_+^T), \qquad
\bW_2 = - \kappa^2 \left( \frac{1}{2} \, \partial_-(\bSigma_+ +
\bSigma_+^T) + \bSigma_+ \bSigma_+^T \right), \label{63}
\end{eqnarray}
where
\[
\Sigma_- = \Gamma^{-1} \partial_- \Gamma, \qquad \bSigma_+ =
\partial_+ \Gamma \Gamma^{-1}.
\]
Here we have
\[
W_1^T = - W_1, \qquad \bW_1^T = - \bW_1, \qquad W_2^T = W_2, \qquad
\bW_2^T = \bW_2.
\]
Therefore, in this case there are $2 n^2$ independent characteristic
integrals.

It is convenient to define the scalar product in
$\mathfrak{sp}_n(\bbR)$ as
\[
B(x, y) = \frac{1}{2} \, \tr(xy).
\]
Taking into account the relations (\ref{60}), (\ref{61}), (\ref{62}) 
and (\ref{63}) we come to the following expressions for the nonzero
components of the conformally improved energy-momentum tensor
\begin{equation}
T'_{--} = \frac{1}{2\kappa} \, \tr \left[ W_1^2 - 2 \, W_2 \right],
\qquad T'_{++} = \frac{1}{2\kappa} \, \tr \left[ \bW_1^2 - 2 \, \bW_2
\right]. \label{68}
\end{equation}

Proceed now to the Hamiltonian formalism. It is clear that the
matrices
\[
E_i{}^j = \left( \begin{array}{cc}
e_i{}^j & 0 \\
0 & - (e_i{}^j)^T
\end{array} \right), \qquad i, j = 1, \ldots, n,
\]
form a basis of the Lie algebra $\gothg_0$ and one has
\[
B(E_i{}^j, E_k{}^l) = \delta_i^l \, \delta_k^j.
\]
Using the equality
\begin{equation}
e_i{}^j \otimes e_j{}^i = (e_i{}^j)^T \otimes (e_j{}^i)^T, \label{66}
\end{equation}
we see that the element $C \in \gothg_0 \otimes \gothg_0$ is again
given by the formula (\ref{64}). Let us use for the mapping
$\gamma$ the parametrization (\ref{65}) and for the quantities $j$ 
and $\bar \jmath$ the parametrizations
\[
j = \left( \begin{array}{cc}
\mathcal J & 0 \\
0 & -\mathcal J^T
\end{array} \right), \qquad 
\bar \jmath = \left( \begin{array}{cc}
\bar {\mathcal J} & 0 \\
0 & - \bar {\mathcal J}^T
\end{array} \right),
\]
where the functions $\mathcal J$ and $\bar {\mathcal J}$
take values in $\mathfrak{gl}_n(\bbR)$. Now the relations
(\ref{27})--(\ref{29}) give
\begin{eqnarray*}
& \{\mathcal J(x) \cmotms \mathcal J(x') \} 
= {} - [C_n, I_n \otimes \mathcal J(x)] \, \delta(x - x') 
- 2 \kappa \, C_n \, \delta'(x - x'),
\\[.5em]
& \{\bar \mathcal J(x) \cmotms \bar \mathcal J(x') \} 
= [C_n, I_n \otimes \bar \mathcal J(x)] \, \delta(x - x') 
+ 2 \kappa \, C_n \, \delta'(x - x'), \\[.5em]
& \{\mathcal J(x) \cmotms \bar \mathcal J(x') \} = 0,
\end{eqnarray*}
and the relations (\ref{30}) and (\ref{31}) imply
\begin{eqnarray*}
\{\Gamma(x) \cmotms \mathcal J(x')\} &=& {} - (\Gamma(x) \otimes I_n)
\, C_n \, \delta(x - x'), \\
\{\Gamma(x) \cmotms \bar \mathcal J(x')\} &=& {} - C_n \, 
(\Gamma(x) \otimes I_n) \, \delta(x - x').
\end{eqnarray*}

The Hamiltonian counterparts of the characteristic integrals $W_1$
and $W_2$ are
\begin{eqnarray*}
& \cW_1 = \mathcal J - \mathcal J^T, \\[.3em]
& \cW_2 = - \mathcal J \mathcal J^T - \kappa \, 
(\partial_x \mathcal J
+ \partial_x \mathcal J^T) + \kappa^2 \Gamma^{-1} (\Gamma^T)^{-1}.
\end{eqnarray*}
To find the Poisson brackets for the characteristic integrals $\cW_1$
and $\cW_2$ we need to know the Poisson brackets between $\Gamma(x)$,
$\mathcal J(x)$ and $\Gamma^T(x)$, $\mathcal J^T(x)$. They can be
found in the following way. Let $\sigma$ be a linear operator on
$\Mat_n(\bbR)$ acting as
\[
\sigma(a) = a^T.
\]
{}From (\ref{A.32}) it follows that 
\[
\{\mathcal J(x) \cmotms \mathcal J^T(x')\} = (\id_{\Mat_n(\bbR)}
\otimes \, \sigma) (\{\mathcal J(x) \cmotms \mathcal J(x')\}),
\]
hence, we have
\[
\{\mathcal J(x) \cmotms \mathcal J^T(x')\} = [\widetilde C_n, I_n
\otimes \mathcal J^T(x)] \, \delta(x - x') - 2 \kappa \, \widetilde
C_n \, \delta'(x - x'),
\]
where
\[
\widetilde C_n = (\id_{\Mat_n(\bbR)} \otimes \, \sigma)(C_n) =
e_i{}^j \otimes(e_j{}^i)^T.
\]
Note that the element $\widetilde C_n$ can also be defined as
\[
\widetilde C_n = (\sigma \otimes \id_{\Mat_n(\bbR)})(C_n) =
(e_i{}^j)^T \otimes e_j{}^i.
\]
Using this relation and the equality (\ref{66}), we obtain
\begin{eqnarray*}
&& \{\mathcal J^T(x) \cmotms \mathcal J(x')\} = {} - [\widetilde C_n,
I_n \otimes \mathcal J(x)] \, \delta(x - x') - 2 \kappa \, \widetilde
C_n \, \delta'(x - x'), \\
&& \{\mathcal J^T(x) \cmotms \mathcal J^T(x')\} = [C_n, I_n \otimes
\mathcal J^T(x)] \, \delta(x - x') - 2 \kappa \, C_n \, 
\delta'(x - x').
\end{eqnarray*}
In a similar way we come to the expressions
\begin{eqnarray*}
&& \{\Gamma(x) \cmotms \mathcal J^T(x')\} = {} - (\Gamma(x) \otimes
I_n) \, \widetilde C_n \, \delta(x - x'), \\
&& \{\Gamma^T(x) \cmotms \mathcal J(x')\} = {} - \widetilde C_n \,
(\Gamma^T(x) \otimes I_n) \, \delta(x - x'), \\
&& \{\Gamma^T(x) \cmotms \mathcal J^T(x')\} = {} - C_n \,
(\Gamma^T(x) \otimes I_n) \, \delta(x - x').
\end{eqnarray*}

It can be verified that the element $\widetilde C_n$ satisfies the
relations
\begin{eqnarray*}
& \widetilde C_n(a \otimes b) = \widetilde C_n (I_n \otimes a^T b) =
\widetilde C_n(b^T a \otimes I_n), \\[.3em]
& (a \otimes b) \widetilde C_n = (I_n \otimes b a^T) \widetilde C_n =
(a b^T \otimes I_n) \widetilde C_n
\end{eqnarray*}
which are used in obtaining the Poisson brackets for $\cW_1$ and
$\cW_2$. These Poisson brackets have the form
\begin{eqnarray*}
\{\cW_1(x) && {} \cmotms \cW_1(x')\} 
= {}- [C_n - \widetilde C_n, I_n \otimes \cW_1(x)] \, \delta (x - x') 
- 4 \kappa \, (C_n - \widetilde C_n) \, \delta'(x - x'), \\
\{\cW_1(x) && {} \cmotms \cW_2(x') \} = {} - [C_n - \widetilde C_n,
I_n \otimes \cW_2(x)] \, \delta(x - x') \\* 
&& \hspace{10em} {} - \kappa \, [C_n - \widetilde C_n, I_n \otimes
\cW_1(x')]_+ \, \delta'(x - x'), \\[.3em]
\{\cW_2(x) && {} \cmotms \cW_2(x')\} 
= (I_n \otimes \cW_2(x)) (C_n + \widetilde C_n) (I_n \otimes \cW_1(x))
\, \delta(x - x')  \\
&& {} - (I_n \otimes \cW_1(x))(C_n + \widetilde C_n) 
(I_n \otimes \cW_2(x)) \, \delta(x - x') \\[.3em]
&& {} - \frac{\kappa^2}{2} \, [C_n + \widetilde C_n, I_n \otimes
\partial_x^2 \mathcal W_1(x)] \, \delta(x - x') \\[.3em]
&& {} + \kappa \, [C_n + \widetilde C_n, I_n \otimes (\cW_2(x) +
\mathcal W_2(x'))]_+ \, \delta'(x - x') \\[.5em]
&& {} - \kappa \, (I_n \otimes \cW_1(x)) (C_n + \widetilde C_n) (I_n
\otimes \cW_1(x)) \, \delta'(x - x') \\[.3em]
&& {} - \kappa \, (I_n \otimes \cW_1(x')) (C_n + \widetilde C_n) (I_n
\otimes \cW_1(x')) \, \delta'(x - x') \\[.3em]
&& {} + \frac{3 \kappa^2}{2} \, [C_n + \widetilde C_n, I_n \otimes
(\cW_1(x) + \cW_1(x'))] \, \delta''(x - x') + 4 \kappa^3 \, (C_n +
\widetilde C_n) \, \delta'''(x - x').
\end{eqnarray*}
Note that the first two expressions above can be obtained from
the relations (\ref{55}) and (\ref{42}) replacing $C_n$ by $C_n -
\widetilde C_n$. One can also obtain the third expression from
(\ref{43}) replacing $C_n$ by $C_n + \widetilde C_n$. The same
procedure can be used to obtain the Poisson brackets for the
Hamiltonian counterparts of the characteristic integrals $\bW_1$ 
and $\bW_2$, which are of the form
\begin{eqnarray*}
& \bcW_1 = \bar \mathcal J - \bar \mathcal J^T, \\[.3em]
& \bcW_2 = - \bar \mathcal J^T \bar \mathcal J + \kappa \,
(\partial_x \bar \mathcal J + \partial_x \bar \mathcal J^T) 
+ \kappa^2 (\Gamma^T)^{-1} \Gamma^{-1},
\end{eqnarray*}
from the relations (\ref{52})--(\ref{54}).

As follows from (\ref{68}) the Hamiltonian counterparts of 
the nonvanishing components of the energy-momentum tensor are 
\[
\mathcal T'_{--} = \frac{1}{2\kappa} \, \tr \left[ \cW_1^2 - 2 \,
\cW_2 \right], \qquad \mathcal T'_{++} = \frac{1}{2\kappa} \, \tr
\left[ \bcW_1^2 - 2 \, \bcW_2 \right].
\]
The quantities $\mathcal V(x)$ and $\bar \mathcal V(x)$, defined by
the relation (\ref{67}), again give two copies of the Virasoro
algebra:
\begin{eqnarray*}
& \{\mathcal V(x), \mathcal V(x')\} = {} 
- (\mathcal V(x) + \mathcal V(x')) \, \delta'(x - x') 
+ \frac{\kappa}{2} \, n \, \delta'''(x - x'), \\[.3em]
& \{\bar \mathcal V(x), \bar \mathcal V(x')\} = (\bar \mathcal V(x) +
\bar \mathcal V(x')) \, \delta'(x - x') - \frac{\kappa}{2} \, n \,
\delta'''(x - x'), \\[.3em]
& \{ \mathcal V(x), \bar \mathcal V(x') \} = 0.
\end{eqnarray*}
They generate the conformal transformations (\ref{82}). It can be
shown that the characteristic integrals $\cW_1(x)$ and $\bcW_1(x)$
have the conformal weight 1 with respect to $\mathcal V(x)$ and 
$\bar \mathcal V(x)$ respectively. The conformal weight of the 
characteristic integrals $\cW_2(x)$ and $\bcW_2(x)$ with respect 
to $\mathcal V(x)$ and $\bar \mathcal V(x)$, respectively, 
is equal to 2.

We will not write explicit expressions for the infinitesimal symmetry
transformations generated by the characteristic integrals. They are
similar to the transformations described by relations (\ref{73}),
(\ref{74}) and (\ref{75}), (\ref{76}). Note only that the
characteristic integrals $\cW_1$ and $\bcW_1$ generate WZNW-type
symmetry transformations given by (\ref{83}).

\section{Conclusion}

We found the classical $W$-algebras and the corresponding
infinitesimal symmetry transformations for the simplest non-abelian
Toda systems associated with the Lie groups $\mathrm{GL}_{\,
2n}(\bbR)$ and $\mathrm{Sp}_n(\bbR)$. The block matrix structure of
the systems under consideration results in the fact that the
generators of the $W$--algebras appear as matrix-valued quantities.
Actually, it is this fact that gives us a possibility to write the
defining relations in a compact form.

To obtain the $W$-algebras for the case of the Toda systems related to
the Lie group $\mathrm{Sp}_n(\bbR)$ one could also use the fact that
the symplectic group is a subgroup of the general linear group,
implementing the reduced phase space formalism. In such a case, one
should work with the corresponding Dirac bracket. However, the
calculations one should perform along that line of approach turn out
to be more cumbersome than those we have done. So, the more direct
approach to the problem presented here is certainly preferable, at
least from a technical point of view.  

It is worth to note that the generators of the $W$-algebras obtained
in the paper have the conformal spin 1 or 2. Nevertheless, we gain 
nonlinear defining relations. It is not usual for the theory of
$W$-algebras, although it was observed  in the theory of $V$-algebras.
The latter are also extensions of the Virasoro algebra, but they allow
for nonlocal terms in expressions for the Poisson brackets of
generators \cite{Bil94a,Bil94b,Bil95,GSZ98,GSZ99}. It seems that
they can be obtained from the $W$-algebras for non-abelian Toda
systems by imposing the constraints saying that the generators of the
WZNW-type symmetry, $\cW_1$ and $\bcW_1$ in our case, are equal to
zero, but the explicit relationship requires further investigation.
The main problem here is to determine the structure of the reduced
phase space in the case when the group formed by the WZNW-type
symmetry transformations is non-abelian.

The work of A.V.R. was supported in part by the Russian Foundation
for Basic Research under grant \#01--01--00201 and the INTAS under
grant \#00--00561.

\appendix

\section{Coordinates and metrics conventions}

Let $M$ be a two-dimensional orientable Riemannian manifold $M$ with
metric tensor $\eta$ of index 1. Arbitrary coordinates on $M$ are
denoted by $z^i$ and for the partial derivatives we use the notation
$\partial_i = \partial/\partial z^i$. Starting from the local
representation for $\eta$,
\[
\eta = \eta_{ij} \, \dif z^i \otimes \dif z^j,
\]
we define the quantities $\eta^{ij}$ by
\[
\eta^{ik} \eta_{kj} = \delta^i_j,
\]
and denote $\det \| \eta_{ij} \|$ simply by $\eta$.

In the paper we deal with a flat two-dimensional manifold $M$ and
use flat Minkowski coordinates $z^0$, $z^1$, such that the metric
tensor is
\[
\eta = {} - \dif z^0 \otimes \dif z^0 + \dif z^1 \otimes \dif z^1.
\]
The light-front coordinates $z^-$ and $z^+$ are introduced by the
relations
\begin{equation}
z^- = \frac{1}{2} (z^0 - z^1), \qquad z^+ = \frac{1}{2} (z^0 + z^1).
\label{A.34}
\end{equation}
The inverse transformation to the coordinates $z^0$ and $z^1$ is
given by
\[
z^0 = z^- + z^+, \qquad z^1 = {} - z^- + z^+.
\]
Using the light-front coordinates, we obtain for the metric tensor
\begin{equation}
\eta = {} - 2 \, \dif z^- \otimes \dif z^+ 
- 2 \, \dif z^+ \otimes \dif z^-. \label{A.33}
\end{equation}
The connection of partial derivatives is
\begin{eqnarray*}
&\partial_- = \partial_0 - \partial_1, \qquad 
\partial_+ = \partial_0 + \partial_1, \\[.5em]
&\partial_0 = \frac{1}{2} (\partial_- + \partial_+), \qquad
\partial_1 = \frac{1}{2}( - \partial_- + \partial_+).
\end{eqnarray*}

\section{Some information on matrix Lie groups} \label{appa}

Let $G$ be a real matrix Lie group or, in other words, a Lie subgroup
of the Lie group $\mathrm{GL}_N(\bbR)$. Denote by $y^\mu$ some local
coordinates on $G$ and by $g$ the matrix-valued function which
transforms the coordinates $y^\mu(a)$ of the element $a \in G$ into
the element $a$ itself. The left-invariant Maurer--Cartan form
$\theta$ can be written as
\[
\theta = g^{-1}(y) \dif g(y),
\]
where the matrix-valued function $g^{-1}$ is defined by the equality
\[
g^{-1}(y) g(y) = I_N.
\]
It is easy to verify that $\theta$ satisfies the relation
\begin{equation}
\dif \theta + \theta \wedge \theta = 0. \label{A.1}
\end{equation}
Using the basis of 1-forms $\dif y^\mu$, one can write
\begin{equation}
\theta = g^{-1}(y) \partial_\mu g (y) \, \dif y^\mu = \theta_\mu(y)
\dif y^\mu,
\label{A.17}
\end{equation}
where $\theta_\mu(y)$ are matrix-valued functions on $M$. Recall that
the Maurer--Cartan form is a $\gothg$-valued one-form and
therefore the functions $\theta_\mu(y)$ take values in the Lie
algebra $\gothg$ of the Lie group $G$. The relation (\ref{A.1}) is 
equivalent to the equalities
\begin{equation}
\partial_\mu \theta_\nu(y) - \partial_\nu \theta_\mu(y) +
[\theta_\mu(y), \theta_\nu(y)] = 0. \label{A.6}
\end{equation}

Choose some basis $\{e_\alpha\}$ of $\gothg$ and denote by
$f^\alpha{}_{\beta \gamma}$ the corresponding structure constants,
\[
[e_\alpha, e_\beta] = e_\gamma f^\gamma{}_{\alpha \beta}.
\]
Expand the functions $\theta_\mu(y)$ over the basis
$\{e_\alpha\}$,
\begin{equation}
\theta_\mu(y) = e_\alpha \theta^\alpha_\mu(y). \label{A.7}
\end{equation}
This gives the following representation for the left-invariant
Maurer--Cartan form:
\begin{equation}
\theta = e_\alpha \theta^\alpha_\mu(y) \dif y^\mu. \label{A.25}
\end{equation}
The relation (\ref{A.6}) implies that
\begin{equation}
\partial^{}_\mu \theta^\alpha_\nu(y) - \partial^{}_\nu
\theta^\alpha_\mu(y) + f^\alpha{}^{}_{\beta \gamma}
\theta^\beta_\mu(y) \theta^\gamma_\nu(y) = 0. \label{A.2}
\end{equation}
Denote by $X^\mu_\alpha(y)$ the functions satisfying the relation
\begin{equation}
X^\mu_\alpha(y) \theta^\alpha_\nu(y) = \delta^\mu_\nu. \label{A.15}
\end{equation}
Note that the functions $X^\mu_\alpha(y)$ are the components of the
left-invariant vector fields $X_\alpha = X^\mu_\alpha(y)
\partial_\mu$
on $G$. It is not difficult to see that (\ref{A.2}) implies
\begin{equation}
X^\mu_\alpha(y) \, \partial_\mu X^\nu_\beta(y) - X^\mu_\beta(y) \,
\partial_\mu X^\nu_\alpha(y) = X^\nu_\gamma(y) f^\gamma{}_{\alpha
\beta} \label{A.11}
\end{equation}
that, in terms of the vector fields $X_\alpha$, can be written as
\[
[X_\alpha, X_\beta] = X_\gamma f^\gamma{}_{\alpha \beta}.
\]

The right-invariant Maurer--Cartan form
\[
\bar \theta = \dif g (y) g^{-1}(y)
\]
satisfies the relation
\[
\dif \bar \theta - \bar \theta \wedge \bar \theta = 0.
\]
Introducing the local expansion
\begin{equation}
\bar \theta = e_\alpha \bar \theta_\mu^\alpha(y) \dif y^\mu,
\label{A.26}
\end{equation}
we obtain for the functions $\bar \theta_\mu^\alpha(y)$ the following
equalities:
\begin{equation}
\partial_\mu \bar \theta_\nu^\alpha(y) 
- \partial_\nu \bar\theta_\mu^\alpha(y)
- f^\alpha{}_{\beta \gamma} 
\bar \theta_\mu^\beta(y) \bar\theta_\nu^\gamma(y) = 0. \label{A.12}
\end{equation}
The functions $\bX^\mu_\alpha(y)$ defined by
\begin{equation}
\bX^\mu_\alpha(y) \bar \theta^\alpha_\nu(y) = \delta^\mu_\nu
\label{A.16}
\end{equation}
are the components of the right-invariant vector fields $\bX_\alpha
= \bX^\mu_\alpha(y) \partial_\mu$ on $G$. The equalities (\ref{A.12})
imply that
\begin{equation}
\bX^\mu_\alpha(y) \, \partial_\mu \bX^\nu_\beta(y) - \bX^\mu_\beta(y)
\, \partial_\mu \bX^\nu_\alpha(y) = - \bX^\nu_\gamma(y)
f^\gamma{}_{\alpha \beta} \label{A.13}
\end{equation}
that is equivalent to
\[
[\bX_\alpha, \bX_\beta] = - \bX_\gamma f^\gamma{}_{\alpha \beta}.
\]

{}From (\ref{A.17}) and (\ref{A.7}) we obtain
\begin{equation}
X_\alpha g(y) = g(y) \, e_\alpha. \label{A.18}
\end{equation}
This means that the vector fields $X_\alpha$ are generators of right
shifts in the Lie group $G$. Correspondingly, the equality
\begin{equation}
\bX_\alpha g(y) = e_\alpha g(y) \label{A.19}
\end{equation}
tells us that the vector fields $\bX_\alpha$ are generators of left
shifts. Now, using (\ref{A.18}) and (\ref{A.19}), we see that
\[
[X_\alpha, \bX_\beta] \, g(y) = 0
\]
that implies the equality
\[
[X_\alpha, \bX_\beta] = 0.
\]
In terms of the components we present the above equality in the 
form
\[
X^\mu_\alpha(y) \, \partial^{}_\mu \bX^\nu_\beta(y) -
\bX^\mu_\beta(y)
\, \partial^{}_\mu X^\nu_\alpha(y) = 0.
\]

Since we are working with a matrix Lie group $G$, the adjoint
representation of $G$ can be defined by the relation
\[
\Ad (a) x = a x a^{-1},
\]
and the matrix of $\Ad(a)$ with respect to the basis $\{e_\alpha\}$ 
of $\gothg$ is defined by the equality
\[
\Ad (a) e_\alpha = e_\beta \Ad^\beta{}_\alpha(a).
\]
Consider the action of the left invariant vector field $X_\alpha$ 
on the matrix-valued function $g \, e_\beta \, g^{-1}$. Using the
equality
\[
X_\alpha g^{-1}(y) = - g^{-1}(y) X_\alpha g(y) \, g^{-1}(y)
\]
and the relation (\ref{A.18}), we obtain
\[
X_\alpha (g(y) \, e_\beta \, g^{-1}(y)) = g(y) \, e_\gamma \,
g^{-1}(y) f^\gamma{}_{\alpha \beta}.
\]
In terms of the matrix elements we obtain
\[
X_\alpha (\Ad^\gamma{}_\beta(g(y))) = \Ad^\gamma{}_\delta(g(y))
f^\delta{}_{\alpha \beta}.
\]
It is not difficult to show that
\begin{equation}
X_\alpha (\Ad^\gamma{}_\beta(g^{-1}(y)) = - f^\gamma{}_{\alpha
\delta}
\Ad^\delta{}_\beta(g^{-1}(y)). \label{A.27}
\end{equation}

Recall that for any $a \in G$ the operator $\Ad(a)$ is an
automorphism of $\gothg$:
\[
\Ad(a)[x,y] = [\Ad(a)x, \Ad(a)y].
\]
In terms of components this equality takes the form
\[
\Ad^\alpha{}_\delta (a) f^\delta{}_{\beta \gamma} =
f^\alpha{}_{\varepsilon \zeta} \Ad^\varepsilon{}_\beta(a)
\Ad^\zeta{}_\gamma(a).
\]

The right-invariant Maurer--Cartan form and the left-invariant one
are connected by the relation
\[
\bar \theta = g(y) \, \theta \, g^{-1}(y).
\]
Using (\ref{A.25}) and (\ref{A.26}), we obtain the equality
\begin{equation}
\Ad^\alpha{}_\beta (g(y)) = \bar \theta^\alpha_\mu(y) X^\mu_\beta(y).
\label{A.23}
\end{equation}

Suppose that the Lie algebra $\gothg$ is endowed with a nondegenerate
symmetric invariant scalar product. This means that there is given a
bilinear mapping $B: \gothg \times \gothg \to \bbR$, satisfying the
relations
\begin{eqnarray}
& B(x, y) = B(y, x), \label{A.4} \\
& B([x, y], z) = B(x, [y, z]), \label{A.5}
\end{eqnarray}
and the condition that if $B(x, y) = 0$ for all $y \in \gothg$ then
$x = 0$. 

{}From (\ref{A.4}) it follows that for the quantities
\begin{equation}
c_{\alpha \beta} = B(e_\alpha, e_\beta) \label{A.9}
\end{equation}
one has
\[
c_{\alpha \beta} = c_{\beta \alpha},
\]
and the nondegeneracy of the scalar product $B$ implies that the
matrix $\|c_{\alpha \beta}\|$ is invertible. Using the relation
(\ref{A.5}), one can show that the quantities
\[
f_{\alpha \beta \gamma} = c_{\alpha \delta} f^\delta{}_{\beta \gamma}
\]
are totally antisymmetric with respect to the indices $\alpha$,
$\beta$ and $\gamma$. It can also be shown that in the case under
consideration
\[
f^\alpha{}_{\beta \alpha} = 0.
\]
This equality implies that if the Lie group $G$ is connected then it
is unimodular.

Actually we assume in the paper that the scalar product $B$ is
$\Ad$-invariant. It means that for any $a \in G$ and any $x,y \in
\gothg$ one has
\[
B(\Ad(a)x, \Ad(a)y) = B(x, y).
\]
In terms of components and matrix elements we have
\[
c_{\gamma \delta} \Ad^\gamma{}_\alpha(a) \Ad^\delta{}_\beta(a) =
c_{\alpha \beta}.
\]
Using the equality (\ref{A.23}), we obtain
\begin{equation}
c_{\alpha \beta} \, \bar \theta^\alpha_\mu(y) \bar \theta^\beta_\nu(y)
= c_{\alpha \beta} \, \theta^\alpha_\mu(y) \theta^\beta_\nu(y).
\label{A.24}
\end{equation}
The last expressions establish the bi-invariant metric tensor
on the $G$-group manifold, related to the local coordinates $y^\mu$. 

In construction of the action of the WZNW model one uses the
three-form
\begin{equation}
\Theta = \frac{1}{3!} \, B(\theta_\mu(y), [\theta_\nu(y),
\theta_\rho(y)]) \, \dif y^\mu \wedge \dif y^\nu \wedge \dif y^\rho.
\label{A.3}
\end{equation}
Using (\ref{A.6}), one can show that this form is closed, but, in
general, it is not exact. Locally for some two-form
\[
\lambda = \frac{1}{2!} \, \lambda_{\mu \nu}(y) \dif y^\mu \wedge \dif
y^\nu.
\]
we can write
\begin{equation}
\Theta = \dif \lambda. \label{A.8}
\end{equation}
Taking into account (\ref{A.7}), we have
\[
\Theta = \frac{1}{3!} \, f_{\alpha \beta \gamma} \,
\theta_\mu^\alpha(y) \theta_\nu^\beta(y) \theta_\rho^\gamma(y) \dif
y^\mu \wedge \dif y^\nu \wedge \dif y^\rho,
\]
and the relation (\ref{A.8}) implies
\begin{equation}
\partial_\mu \lambda_{\nu \rho}(y) + \partial_\nu \lambda_{\rho
\mu}(y) + \partial_\rho \lambda_{\mu \nu}(y) = f_{\alpha \beta
\gamma}
\, \theta_\mu^\alpha(y) \theta_\nu^\beta(y) \theta_\rho^\gamma(y).
\label{A.14}
\end{equation}

\section{Current algebra} \label{la}

To find the Poisson brackets for $j_\alpha(x)$ write
\begin{eqnarray*}
\{\pi_\mu(x) &+& \kappa \, \lambda_{\mu \rho}(\xi(x)) \, \partial_x
\xi^\rho(x), \pi_\nu(x') + \kappa \, \lambda_{\nu \sigma}(\xi(x')) \,
\partial_x \xi^\sigma(x') \} \\[.5em]
&=& - \kappa \, [\partial_\mu \lambda_{\nu \rho}(\xi(x)) +
\partial_\nu \lambda_{\rho \mu}(\xi(x)) + \partial_\rho \lambda_{\mu
\nu}(\xi(x))] \, \partial_x \xi^\rho(x) \, \delta(x - x').
\end{eqnarray*}
Taking into account equality (\ref{A.14}), we obtain
\begin{eqnarray*}
\{\pi_\mu(x) + \kappa \, \lambda_{\mu \rho}(\xi(x)) \,
&& \partial_x \xi^\rho(x), \pi_\nu(x') + \kappa \, \lambda_{\nu
\sigma}(\xi(x')) \, \partial_x \xi^\sigma(x') \} \\[.5em]
&& = - \kappa \, f_{\alpha \beta \gamma} \, \theta^\alpha_\mu(\xi(x))
\theta^\beta_\nu(\xi(x)) \theta^\gamma_\rho(\xi(x)) \, \partial_x
\xi^\rho(x) \, \delta(x - x').
\end{eqnarray*}
Hence, using the equalities (\ref{A.11}) we find 
\begin{eqnarray}
\{ -X^\mu_\alpha(\xi(x))[\pi_\mu(x) + \kappa \,
\lambda_{\mu\rho}(\xi(x)) \, 
\partial_x \xi^\rho(x)], \, && -
X^\nu_\beta(\xi(x'))[\pi_\nu(x') + \kappa \, \lambda_{\nu
\sigma}(\xi(x')) \, \partial_x \xi^\sigma(x')]\} \nonumber \\[.5em]
= - X^\rho_\gamma(\xi(x))[\pi_\rho(x) && {} + \kappa \,
\lambda_{\rho \sigma}(\xi(x)) \, \partial_x \xi^\sigma(x)] \,
f^\gamma{}_{\alpha \beta} \, \delta(x - x') \nonumber \\[.5em]
&&{} - \kappa \, c_{\gamma \delta} \, \theta^\delta_\sigma(\xi(x)) \,
\partial_x \xi^\sigma(x) \, f^\gamma{}_{\alpha \beta} \, 
\delta(x - x'). \label{A.20}
\end{eqnarray}
It is easy to get convinced that
\begin{eqnarray*}
\{-X^\mu_\alpha && (\xi(x))[\pi_\mu(x) + \kappa \, \lambda_{\mu
\rho}(\xi(x)) \, \partial_x \xi^\rho(x)], \, \kappa \, c_{\beta
\delta} \, \theta^\delta_\sigma(\xi(x')) \,
\partial_x \xi^\sigma(x')\} \\[.5em]
&&= \kappa \, X^\mu_\alpha(\xi(x)) \, c_{\beta \delta}
[\partial_\mu \theta^\delta_\sigma(\xi(x)) - \partial_\sigma
\theta^\delta_\mu(\xi(x))] \partial_x \xi^\sigma(x)\,
\delta(x - x') - \kappa \, c_{\alpha \beta} \, \delta'(x - x').
\end{eqnarray*}
Using the relation (\ref{A.2}), we come to
\begin{eqnarray}
\{-X^\mu_\alpha(\xi(x))[\pi_\mu(x) &+& \kappa \, \lambda_{\mu
\rho}(\xi(x)) \, \partial_x \xi^\rho(x)], \, \kappa \, c_{\beta
\delta} \, \theta^\delta_\sigma(\xi(x')) \, \partial_x
\xi^\sigma(x')\} \nonumber \\[.5em]
&=& \kappa \, c_{\gamma \delta} \, \theta^\delta_\sigma(\xi(x))
\, \partial_x \xi^\sigma(x) \, f^\gamma{}_{\alpha \beta} \,
\delta(x - x') - \kappa \, c_{\alpha \beta} \, \delta'(x - x').
\label{A.21}
\end{eqnarray}
Similarly we obtain
\begin{eqnarray}
\{\kappa \, c_{\alpha \gamma} \, \theta^\gamma_\rho(\xi(x)) \,
&& \partial_x \xi^\rho(x), \, -X^\nu_\beta(\xi(x'))[\pi_\nu(x') +
\kappa \, \lambda_{\nu \sigma}(\xi(x')) \, \partial_x
\xi^\sigma(x')]\} \nonumber \\[.5em]
&&= \kappa \, c_{\gamma \delta} \, \theta^\delta_\sigma(\xi(x)) \,
\partial_x \xi^\sigma(x) \, f^\gamma{}_{\alpha \beta} \, 
\delta(x - x') - \kappa \, c_{\alpha \beta} \, \delta'(x - x'). 
\label{A.22}
\end{eqnarray}
Finally, the equalities (\ref{A.20}), (\ref{A.21}) and (\ref{A.22}) 
give the relation (\ref{21}).

To find the expression for the Poisson brackets of $\bar
\jmath_\alpha$ we write the equality
\begin{equation}
\bar \jmath_\alpha = - \left( X^\mu_\beta(\xi) [\pi_\mu + \kappa
\, \lambda_{\mu \rho}(\xi) \, \partial_x \xi^{\rho}] 
+ \kappa \, c_{\beta\gamma} \, \theta^\gamma_\mu (\xi) \, 
\partial_x \xi^{\mu} \right) \Ad^\beta{}_\alpha(g^{-1}) \label{A.35}
\end{equation}
which follows from (\ref{A.23}) and (\ref{A.24}). Now using
(\ref{A.20})--(\ref{A.22}) and (\ref{A.27}) we obtain (\ref{22}). 
In a similar way we arrive at the relation (\ref{23}).

\section{Algebra-valued functions on a phase space}
\label{av}

Let $A_1$ and $A_2$ be two unital algebras with the units $1_{A_1}$
and $1_{A_2}$. The Poisson bracket of an $A_1$-valued function $F$
and an $A_2$-valued function $G$ on a symplectic manifold $M$ is
defined
in the following way. Choose some basis $\{e_\alpha\}$ of $A_1$ and
some basis $\{f_i\}$ of $A_2$. Expand the functions $F$ and $G$ over
the bases,
\[
F = e_\alpha F^\alpha, \qquad G = f_i \, G^i.
\]
Then the Poisson bracket of $F$ and $G$ is defined as an $A_1 \otimes
A_2$-valued function
\[
\{F \cmotms G \} = e_\alpha \otimes f_i \, \{F^\alpha, G^i\}.
\]
The Poisson bracket $\{F \cmotms G\}$ does not depend on the choice 
of the bases $\{e_\alpha\}$ and $\{f_i\}$.

Introducing the linear mapping $P$ from $A_1 \otimes A_2$ to $A_2
\otimes A_1$ defined by the relation
\[
P(a \otimes b) = b \otimes  a,
\]
we can reformulate the usual properties of the Poisson bracket as
follows: 
\begin{eqnarray*}
& \{F \cmotms G\} = - P \circ \{G \cmotms F\}, \\
& \{F \cmotms G H\} = \{F \cmotms G\} (1_{A_1} \otimes H) + (1_{A_1}
\otimes G) \{F \cmotms H\}, \\
& \{F G \cmotms H\} = (F \otimes 1_{A_2}) \{G \cmotms H\} + \{F
\cmotms H\} (G \otimes 1_{A_2}).
\end{eqnarray*}
It is clear that $P$ is an isomorphism of the algebras $A_1 \otimes
A_2$ and $A_2 \otimes A_1$. The Jacobi identity for the usual Poisson
bracket implies
\[
P_{13} \circ \{F \cmotms \{G \cmotms H \}\} + P_{23} \circ \{H
\cmotms
\{F \cmotms G\}\}) + P_{12} \circ \{G \cmotms \{H \cmotms F\}\} = 0.
\]
Here $P_{13}$ is the linear mapping from $A_1 \otimes A_2 \otimes
A_3$
to $A_3 \otimes A_2 \otimes A_1$ permuting the first and the third
factors:
\[
P_{13}(a \otimes b \otimes c) = c \otimes b \otimes a.
\]
The linear mappings $P_{12}$ and $P_{23}$ are defined 
analogously.

Let $\sigma_1$ be a linear mapping from an algebra $A_1$ to an
algebra $B_1$, and $\sigma_2$ be a linear mapping from an algebra 
$A_2$ to an algebra $B_2$. It can be easily shown that
\begin{equation}
\{\sigma_1 \circ F \cmotms G\} = (\sigma_1 \otimes \id_{A_2})(\{F
\cmotms
G\}), \quad \{F \cmotms \sigma_2 \circ G\} = (\id_{A_1} \otimes \,
\sigma_2)(\{F \cmotms G\}). \label{A.32}
\end{equation}

\section{W-algebra calculations} \label{wc}

To obtain the expression for the Poisson bracket of $\cW_1$ and
$\cW_2$ we find first
\begin{eqnarray}
\{\mathcal J^{(1)}(x) && + \mathcal J^{(2)}(x) \cmotms \mathcal
J^{(1)}(x') \mathcal J^{(2)}(x') \} = {} -[C_n, I_n \otimes \mathcal
J^{(1)}(x) \mathcal J^{(2)}(x)] \, \delta(x - x') \nonumber \\
&& {} - 2 \kappa \, C_n \, (\mathcal J^{(1)}(x') \otimes I_n) \,
\delta'(x - x') - 2 \kappa \, (\mathcal J^{(2)}(x') \otimes I_n) \,
C_n \, \delta'(x - x'), \label{A.28} \\[.5em]
\{ \mathcal J^{(1)}(x) && + \mathcal J^{(2)}(x) \cmotms
-\kappa(\partial_x \mathcal J^{(1)}(x') - \partial_x \mathcal
J^{(2)}(x')) \} \nonumber \\
&& = {} - [C_n, I_n \otimes - \kappa \, (\partial_x \mathcal
J^{(1)}(x) - \partial_x \mathcal J^{(2)}(x))] \, \delta (x - x')
\nonumber \\[.5em]
&& \hspace{7em} {} - \kappa \, [C_n, I_n \otimes (\mathcal
J^{(1)}(x')
- \mathcal J^{(2)}(x'))] \, \delta'(x - x'). \label{A.29}
\end{eqnarray}
We also need the Poisson brackets of $\mathcal J^{(r)}$ with
$\Gamma^{(r)-1}$. To find them note that relation (\ref{40}) implies
\[
\{ \mathcal J^{(r)}(x) \cmotms \Gamma^{(s)}(x') \} = (I_n \otimes
\Gamma^{(r)}(x)) \, C_n \, \delta(x - x') \, \delta^{rs}.
\]
Writing now the relation
\begin{eqnarray*}
\{ \mathcal J^{(r)}(x) \cmotms \Gamma^{(s)}(x') \Gamma^{(s)-1}(x') \}
= (I_n \otimes && \Gamma^{(r)}(x)) \, C_n \, (I_n \otimes
\Gamma^{(s)-1}(x')) \, \delta(x - x') \, \delta^{rs} \\
&& {} + (I_n \otimes \Gamma^{(s)}(x')) 
\{ \mathcal J^{(r)}(x) \cmotms \Gamma^{(s)-1}(x') \} = 0,
\end{eqnarray*}
we obtain
\[
\{ \mathcal J^{(r)}(x) \cmotms \Gamma^{(s)-1}(x') \} = {} - C_n \,
(I_n \otimes \Gamma^{(r)-1}(x)) \, \delta(x - x') \, \delta^{rs}.
\]
Using this equality, we come to
\begin{eqnarray}
\{ \mathcal J^{(1)}(x) + \mathcal J^{(2)}(x) \cmotms \kappa^2
&& \Gamma^{(1)-1}(x') \Gamma^{(2)}(x') \} \nonumber \\
&& = {} - [C_n, I_n \otimes \kappa^2 \Gamma^{(1)-1}(x)
\Gamma^{(2)}(x)] \, \delta(x - x'). \label{A.30}
\end{eqnarray}
Collecting the equalities (\ref{A.28}), (\ref{A.29}) and (\ref{A.30}), 
we obtain the relation (\ref{42}).

The calculation of the Poisson bracket for ${\mathcal W}_2$ is more 
complicated. The main formulas used here are
\begin{eqnarray*}
\{ && \cJ{(1)}(x) \cJ{(2)}(x) \cmotms \cJ{(1)}(x') \cJ{(2)}(x') \}
\\*
&& \hspace{1em} {} = {} - [C_n, (\cJ{(1)}(x) + \cJ{(2)}(x)) \otimes
\cJ{(1)}(x) \cJ{(2)}(x) ] \, \delta(x - x') \\*[.5em]
&& \hspace{2.4em} {} - 2 \kappa \, C_n \, (\cJ{(1)}(x') \otimes
\cJ{(1)}(x)) \, \delta'(x - x') - 2 \kappa \, C_n \, (\cJ{(2)}(x)
\otimes \cJ{(2)}(x')) \, \delta'(x - x'), \\[.5em]
\{ && \cJ{(1)}(x) \cJ{(2)}(x) \cmotms -\kappa\, (\partial_x
\cJ{(1)}(x') - \partial_x \cJ{(2)}(x')) \} \\*
&& \hspace{1em} = \kappa \, [C_n, I_n \otimes \cJ{(1)}(x)
\cJ{(2)}(x)]_+ \, \delta'(x - x') - 2 \kappa \, C_n \, (\cJ{(2)}(x)
\otimes \cJ{(1)}(x)) \, \delta'(x - x') \\*[.5em]
&& \hspace{2.4em} {} + 2 \kappa^2 \, C_n \, (I_n \otimes \cJ{(1)}(x))
\, \delta''(x - x') - 2 \kappa^2 \, C_n \, (I_n \otimes \cJ{(2)}(x))
\, \delta''(x - x'), \\[.5em]
\{ && \cJ{(1)}(x) \cJ{(2)}(x) \cmotms \kappa^2 \, \Gamma^{(1)-1}(x')
\Gamma^{(2)}(x') \} \\*
&& \hspace{5em} {} = \kappa^2 \, (\cJ{(1)}(x) \otimes
\Gamma^{(1)-1}(x) \Gamma^{(2)}(x)) \, C_n \, \delta(x - x') \\*[.5em]
&& \hspace{10em} {} - \kappa^2 \, C_n \, (\cJ{(2)}(x)
\otimes \Gamma^{(1)-1}(x) \Gamma^{(2)}(x)) \, 
\delta(x - x'),\\[.5em]
\{ && -\kappa\, (\partial_x \cJ{(1)}(x) - \partial_x \cJ{(2)}(x))
\cmotms -\kappa\, (\partial_x \cJ{(1)}(x')
- \partial_x \cJ{(2)}(x')) \} \\*
&& \hspace{1em} = \kappa^2 \, [C_n, I_n \otimes (\partial_x
\cJ{(1)}(x) + \partial_x \cJ{(2)}(x))] \, \delta'(x - x') \\*[.5em]
&& \hspace{2.4em} {} + \kappa^2 \, [C_n, I_n \otimes (\cJ{(1)}(x) +
\cJ{(2)}(x))] \, \delta''(x - x') + 4 \kappa^3 \, C_n \, 
\delta'''(x - x'), \\[.5em]
\{ && -\kappa\, (\partial_x \cJ{(1)}(x) - \partial_x \cJ{(2)}(x))
\cmotms \kappa^2 \Gamma^{(1)-1}(x') \Gamma^{(2)}(x') \} \\
&& \hspace{10em} = \kappa^3 \, [C_n, I_n \otimes \Gamma^{(1)-1}(x')
\Gamma^{(2)}(x')]_+ \, \delta'(x - x').
\end{eqnarray*}
Using these equalities, after some rearrangement of terms we come to
the relation (\ref{43}).

The relations (\ref{52})--(\ref{54}) can be proven in the same way.

Now we will show that
\begin{equation}
\{\cW_r(x) \cmotms \bcW_s(x')\} = 0. \label{A.31}
\end{equation}
For $r=s=1$ this is a direct consequence of the equality (\ref{39}). 
For the cases $r=1$, $s=2$ and $r=2$, $s=1$ we come to (\ref{A.31}) 
through the relations
\begin{eqnarray*}
& \{\cJ{(1)}(x) + \cJ{(2)}(x) \cmotms \kappa^2 \, \Gamma^{(2)}(x')
\Gamma^{(1)-1}(x') \} = 0, \\
& \{ \kappa^2 \, \Gamma^{(1)-1}(x) \Gamma^{(2)}(x) \cmotms
\bcJ{(1)}(x') + \bcJ{(2)}(x') \} = 0.
\end{eqnarray*}
Finally, using the equalities
\begin{eqnarray*}
\{&& \cJ{(1)}(x) \cJ{(2)}(x) \cmotms \kappa^2 \, \Gamma^{(2)}(x')
\Gamma^{(1)-1}(x') \} \\*
&& \hspace{4em} {} = \kappa^2 \, (\cJ{(1)}(x) \Gamma^{(1)-1}(x)
\otimes \Gamma^{(2)}(x) - \Gamma^{(1)-1}(x) \otimes \Gamma^{(2)}(x)
\cJ{(2)}(x)) \, \delta(x - x'), \\[.5em]
\{&& \kappa^2 \, \Gamma^{(1)-1}(x) \Gamma^{(2)}(x) \cmotms
\bcJ{(2)}(x') \bcJ{(1)}(x')\} \\
&& \hspace{4em} {} = {} - \kappa^2 \, (\Gamma^{(1)-1}(x) \bcJ{(1)}(x)
\otimes \Gamma^{(2)}(x) - \Gamma^{(1)-1}(x) \otimes \bcJ{(2)}(x)
\Gamma^{(2)}(x)) \, \delta(x - x'), \\[.5em]
\{&& - \kappa \, (\partial_x \cJ{(1)}(x) - \partial_x \cJ{(2)}(x))
\cmotms \kappa^2 \, \Gamma^{(2)}(x') \Gamma^{(1)-1}(x')\} \\
&& \hspace{4em} {} + \{ \kappa^2 \Gamma^{(1)-1}(x) \Gamma^{(2)}(x)
\cmotms \kappa \, (\partial_x \bcJ{(1)}(x') - \partial_x
\bcJ{(2)}(x')
\} \\
&& \hspace{1.5em} {} = 2 \kappa^3 ({} - \Gamma^{(1)-1}(x) \partial_x
\Gamma^{(1)}(x) \Gamma^{(1)-1}(x) \otimes \Gamma^{(2)}(x) +
\Gamma^{(1)-1}(x) \otimes \partial_x \Gamma^{(2)}(x)) \delta(x - x'),
\end{eqnarray*}
and taking into account the identity
\[
\cJ{(r)} = \Gamma^{(r)-1} \bcJ{(r)} \Gamma^{(r)} + 2\kappa \,
\Gamma^{(r)-1} \partial_x \Gamma^{(r)},
\]
we see that (\ref{A.31}) is valid for the case $r=s=2$ as well.


\small

\begin{thebibliography}{**}

\bibitem{LSa92}
A. N. Leznov and M. V. Saveliev, 
{\em Group-theoretical Methods for Integration of Nonlinear Dynamical
Systems\/} 
(Birkhauser, Basel, 1992).

\bibitem{RSa97}
A. V. Razumov and M. V. Saveliev,
{\em Lie Algebras, Geometry, and Toda-type Systems\/}
(Cambridge University Press, Cambridge, 1997).

\bibitem{Lez95}
A.~N.~Leznov,
The internal symmetry group and methods of field theory for
integrating exactly soluble dynamic systems,
In: {\em Group Theoretical Methods in Physics\/}, Proc. of the 1982
Zvenigorod seminar (New York, Harwood, 1985), 443--457.

\bibitem{GSa95}
J.--L.~Gervais and M.~V.~Saveliev,
Higher grading generalizations of the Toda systems,
Nucl. Phys. {\bf B453} (1995) 449--476;
{\tt arXiv:hep-th/9505047}.
%%CITATION = HEP-TH 9505047;%%

\bibitem{RSa97a}
A.~V.~Razumov and M.~V.~Saveliev,
On some class of multidimensional nonlinear integrable systems,
In: {\it Second International Sakharov Conference in Physics\/}, eds.
I.~M.~Dremin and A.~M.~Semikhatov (World Scientific, Singapore, 1997)
547--551;
{\tt arXiv:hep-th/9607017}.
%%CITATION = HEP-TH 9607017;%%

\bibitem{RSa97b}
A.~V.~Razumov and M.~V.~Saveliev,
Multi-dimensional Toda-type systems,
Theor. Math. Phys. {\bf 112} (1997) 999--1022; 
{\tt arXiv:hep-th/9609031}.
%%CITATION = HEP-TH 9609031;%%

\bibitem{RSa97c}
A.~V.~Razumov and M.~V.~Saveliev,
Maximally non-abelian Toda systems,
Nucl. Phys. {\bf B494} (1997) 657--686;
{\tt arXiv:hep-th/9612081}.
%%CITATION = HEP-TH 9612081;%%

\bibitem{RSZ99}
A.~V.~Razumov, M.~V.~Saveliev and A.~B.~Zuevsky,
Non-abelian Toda equations associated with classical Lie groups,
In: {\em Symmetries and Integrable Systems\/}, Proc. of the
Seminar, ed. A. N. Sissakian (JINR, Dubna, 1999) 190--203; 
{\tt arXiv:math-ph/9909008}.

\bibitem{EGR97}
P.~Etingof, I.~Gelfand and V.~Retakh,
Factorization of differential operators, quasideterminants, and
non-Abelian Toda field equations,
Math. Res. Lett. {\bf 4} (1997) 413--425;
{\tt arXiv:q-alg/9701008}.

\bibitem{EGR98}
P.~Etingof, I.~Gelfand and V.~Retakh,
Non-Abelian integrable systems, quasideterminants and Marchenko
lemma,
Math. Res. Lett. {\bf 5} (1998) 1--12;
{\tt arXiv:q-alg/9707017}.

\bibitem{Lez98}
A.~N.~Leznov,
The exactly integrable systems connected with semisimple 
algebras of the second rank $A_2$, $B_2$, $C_2$, $G_2$,
{\tt arXiv:math-ph/9809012}.

\bibitem{Lez99}
A.~N.~Leznov,
Graded Lie algebras, representation theory, integrable mappings 
and systems. Non-abelian case,
Nucl. Phys. {\bf B543} (1999) 652--672;
{\tt arXiv:math-ph/9810006}.

\bibitem{Nov82}
S.~P.~Novikov,
The Hamiltonian formalism and a multi-valued analogue of Morse 
theory,
Russian Math. Surveys {\bf 37:5} (1982) 1--56. 
%%CITATION =  %%

\bibitem{Wit84}
E.~Witten,
Nonabelian bosonization in two dimensions,
Commun. Math. Phys.  {\bf 92} (1984) 455--472.
%%CITATION = CMPHA,92,455;%%

\bibitem{Zam85}
A.~B.~Zamolodchikov,
Infinite additional symmetries in two-dimensional conformal quantum
field theory,
Theor. Math. Phys. {\bf 65} (1985) 1205--1213.
%%CITATION = TMPHA,65,1205;%%

\bibitem{BSc93}
P.~Bouwknegt and K.~Schoutens,
$\mathcal W$ symmetry in conformal field theory,
Phys. Rep.  {\bf 223} (1993) 183--276;
{\tt arXiv:hep-th/9210010}.
%%CITATION = HEP-TH 9210010;%%

\bibitem{BFORFW90}
J.~Balog, L.~Feh\'er, L.~O'Raifeartaigh, P.~Forgacs and A.~Wipf,
Toda theory and $\mathcal W$-algebra from a gauged WZNW point of view,
Ann. Phys.  {\bf 203} (1990) 76--136.
%%CITATION = APNYA,203,76;%%

\bibitem{FORTW92a}
L.~Feh\'er, L.~O'Raifeartaigh, P.~Ruelle I.~Tsutsui and A.~Wipf,
Generalised Toda theories and $\mathcal W$-algebras associated with
integral gradings,
Ann. Phys. {\bf 213} (1992) 1--20.
%%CITATION = APNYA,213,1;%%

\bibitem{FORTW92b}
L.~Feh\'er, L.~O'Raifeartaigh, P.~Ruelle, I.~Tsutsui and A.~Wipf,
On Hamiltonian reductions of the Wess--Zumino--Novikov--Witten
theories,
Phys.~Rep. {\bf 222} (1992) 1--64.
%%CITATION = PRPLC,222,1;%%

\bibitem{TFe95}
I.~Tsutsui and L.~Feh\'er,
Global aspects of the WZNW reduction to Toda theories,
Prog. Theor. Phys. Suppl.  {\bf 118} (1995) 173--190;
{\tt arXiv:hep-th/9408065}.
%%CITATION = HEP-TH 9408065;%%

\bibitem{Ful96}
T.~F\"ul\"op,
Reduced SL(2,R) WZNW quantum mechanics,
J. Math. Phys. {\bf 37} (1996) 1617--1631;
{\tt arXiv:hep-th/9502145}.
%%CITATION = HEP-TH 9502145;%%

\bibitem{FTs97}
L.~Feh\'er and I.~Tsutsui,
Regularization of Toda lattices by Hamiltonian reduction,
J. Geom. Phys. {\bf 21} (1997) 97--135;
{\tt arXiv:hep-th/9511118}.
%%CITATION = HEP-TH 9511118;%%

\bibitem{KTs96}
H.~Kobayashi and I.~Tsutsui,
Quantum mechanical Liouville model with attractive potential,
Nucl. Phys. {\bf B472} (1996) 409--426;
{\tt arXiv:hep-th/9601111}.
%%CITATION = HEP-TH 9601111;%%

\bibitem{RYa97}
A.~V.~Razumov and V.~I.~Yasnov,
Hamiltonian reduction of free particle motion on the group
SL$(2,\bbR)$,
Theor.\ Math.\ Phys.\  {\bf 110} (1997) 119--128;
{\tt arXiv:hep-th/9609030}.
%%CITATION = HEP-TH 9609030;%%

\bibitem{BFP98}
J.~Balog, L.~Feh\'er and L.~Palla,
Coadjoint orbits of the Virasoro algebra and the global Liouville
equation,
Int. J. Mod. Phys. A {\bf 13} (1998) 315--362;
{\tt arXiv:hep-th/9703045}.
%%CITATION = HEP-TH 9703045;%%

\bibitem{BNVW99}
Z.~Bajnok, D.~Nogradi, D.~Varga and F.~Wagner,
Geometric quantization of the global Liouville mechanics,
J. Phys. A {\bf 32} (1999) 7477--7481;
{\tt arXiv:hep-th/9906186}.
%%CITATION = HEP-TH 9906186;%%

\bibitem{RSa94}
A.~V.~Razumov and M.~V.~Saveliev,
Differential geometry of Toda systems,
Commun. Anal. Geom.  {\bf 2} (1994) 461--511;
{\tt arXiv:hep-th/9311167}.
%%CITATION = HEP-TH 9311167;%%

\bibitem{OWi90}
L.~O'Raifeartaigh and A.~Wipf,
Conformally reduced WZNW theories and two-dimensional gravity,
Phys. Lett. {\bf B251} (1990) 361--368.
%%CITATION = PHLTA,B251,361;%%

\bibitem{LSa89}
A.~N.~Leznov and M.~V.~Saveliev,
Exactly and completely integrable nonlinear dynamical systems,
Acta Appl. Math. {\bf 16} (1989) 1--74.
%%CITATION = AAMAD,16,1;%%

\bibitem{BGe89a}
A.~Bilal and J.--L.~Gervais,
Extended $c = \infty$ conformal systems from classical Toda field
theories,
Nucl. Phys. {\bf B314} (1989) 646--686.
%%CITATION = NUPHA,B314,646;%%

\bibitem{BGe89b}
A.~Bilal and J.--L.~Gervais,
Systematic construction of conformal theories with higher spin
Virasoro symmetries,
Nucl. Phys. B {\bf B318} (1989) 579--642.
%%CITATION = NUPHA,B318,579;%%

\bibitem{DSo84}
V.~G.~Drinfeld and V.~V.~Sokolov,
Lie algebras and equations of Korteweg--de Vries type,
J.~Sov.~Math.  {\bf 30} (1984) 1975--2036.
%%CITATION = JOSMA,30,1975;%%

\bibitem{Bow89}
P.~Bowcock,
Canonical quantization of the gauged Wess--Zumino model,
Nucl. Phys. {\bf B316} (1989) 80--100.
%%CITATION = NUPHA,B316,80;%%

\bibitem{Sug68}
H.~Sugawara,
A field theory of currents,
Phys. Rev. {\bf 170} (1968) 1659--1662.
%%CITATION = PHRVA,170,1659;%%

\bibitem{Som68}
C.~M.~Sommerfield,
Currents as dynamical variables,
Phys. Rev. {\bf 176} (1968) 2019--2025.
%%CITATION = PHRVA,176,2019;%%

\bibitem{BTV91}
F.A. Bais, T. Tjin and P. van Driel,
Covariantly coupled chiral algebras, 
Nucl. Phys. {\bf B 357} (1991) 632--654.
%%CITATION = NUPHA,B357,632;%%

\bibitem{Bil94a}
A.~Bilal,
Nonabelian Toda theory: A completely integrable model for strings on
a black hole background,
Nucl. Phys. {\bf B422} (1994) 258--290;
{\tt arXiv:hep-th/9312108}.
%%CITATION = HEP-TH 9312108;%%

\bibitem{Bil94b}
A.~Bilal,
Multi-component KdV hierarchy, V-algebra and non-abelian Toda theory,
Lett. Math. Phys.  {\bf 32} (1994) 103--120;
{\tt arXiv:hep-th/9401167}.
%%CITATION = HEP-TH 9401167;%%

\bibitem{Bil95}
A.~Bilal,
Nonlocal matrix generalizations of $W$-algebras,
Commun. Math. Phys.  {\bf 170} (1995) 117--150;
{\tt arXiv:hep-th/9403197}.
%%CITATION = HEP-TH 9403197;%%

\bibitem{GSZ98}
J.~F.~Gomes, G.~M.~Sotkov and A.~H.~Zimerman,
$SU(2,R)_q$ symmetries of non-abelian Toda theories,
Phys. Lett. {\bf B435} (1998) 49--60;
{\tt arXiv:hep-th/9803122}.
%%CITATION = HEP-TH 9803122;%%

\bibitem{GSZ99}
J.~F.~Gomes, G.~M.~Sotkov and A.~H.~Zimerman,
Nonabelian Toda theories from parafermionic reductions of the WZW
model,
Ann. Phys. {\bf 274} (1999) 289--362.
%%CITATION = APNYA,274,289;%%

\end{thebibliography}
\end{document}